\documentclass[12pt]{article}

\usepackage{graphicx}
\begin{document}
\newcommand{\NP}[1]{Nucl.\ Phys.\ {\bf #1}}
\newcommand{\PL}[1]{Phys.\ Lett.\ {\bf #1}}
\newcommand{\PR}[1]{Phys.\ Rev.\ {\bf #1}}
\newcommand{\PRL}[1]{Phys.\ Rev.\ Lett.\ {\bf #1}}
\newcommand{\PREP}[1]{Phys.\ Rep.\ {\bf #1}}
\newcommand{\PTP}[1]{Prog.\ Theor.\ Phys.\ {\bf #1}}
\newcommand{\MPL}[1]{Mod.\ Phys.\ Lett.\ {\bf #1}}

\newcommand{\IJMP}[1]{Int.\ Jour.\ Mod.\ Phys.\ {\bf #1}}
\newcommand{\JHEP}[1]{JHEP\ {\bf #1}}
\begin{titlepage}
\setcounter{page}{0}
~\\
\vspace{10mm}
\begin{center}
{\Large  Causal dynamical triangulation for non-critical open-closed string field theory }

\vspace{10mm}

{\large Hiroshi\ Kawabe\footnote{e-mail address:
kawabe@yonago-k.ac.jp}} \\
{\em
Yonago National College of Technology \\ 
Yonago 683-8502, Japan} \\
\end{center}

\vspace{8mm}
\centerline{{\bf{Abstract}}}
\vspace{5mm}

We extend the 2 dimensional Causal Dynamical Triangulation (CDT) model from the usual model of closed string to the one of open-closed string.
The matrix-vector model describing the loop gas model is modified so as to possess the nature of the CDT, {\it i.e.} the time foliation structure.
Stochastic quantization method produces interactions of loop and line variables similar to those in the non-critical open-closed string field theories.
By taking an appropriate scaling, we realize an extended model of the generalized CDT (GCDT), which keeps the causality in a broad sense.

\end{titlepage}
\newpage
\renewcommand{\thefootnote}{\arabic{footnote}}
\setcounter{footnote}{0}
\section{Introduction}

Over a decade ago we expected matrix models to realize string field theories.
In the dynamical triangulation (DT) formulated by the matrix models, discrete loops on the random surface describe string interactions through the double scaling limit.
In particular, the interaction of a loop with the spin cluster domain wall, the Ishibashi-Kawai (IK-) type interaction, plays an important role in the construction of the non-critical string field theory\cite{IK1,IK2}.
By the stochastic quantization, hermitiam and real symmetric matrix models formulate the orientable and non-orientable string field theories, respectively\cite{JR,Nak}.
An open string propagates and interacts on the 2D surface with boundaries.
The open-closed string field theories are described by matrix-vector models, which have the algebraic structure containing the Virasoro algebra and some current algebra\cite{AJ,Mog,NE}.
The loop gas model describes strings, each of which is located at a point $x$ in the 1D discrete space, interacting with another one only in the same point $x$ or the neighboring points $x\pm 1$\cite{Kos1,KK}.
The matrix-vector model formulation of the loop gas model naturally includes the IK-type interaction\cite{Kos2,Kos3}.
Then, it possesses the similar algebraic structure as above\cite{EKN}.
However, one of the problems in the DT is that the probability of the splitting interaction is too large to realize the string model with stable propagation.
The situation becomes more serious in higher dimensional space-time model.

The causal dynamical triangulation (CDT) is proposed to improve the above problem\cite{AL}. 
It is originally the model only of loop propagation.
While the permission of splitting interaction violates the causality in the strict sense, the prohibition of the merging interaction keeps the violation still soft.
Such a broad sense of causality is adopted to formulate the generalized CDT (GCDT).
This extension changes the propagator with a smooth surface to the one with many projections\cite{ALWZ1}.
Thanks to the diminution of the triangulation by the time foliation structure, it is expected that the propagation becomes stable with moderate quantum correction.
The string field theory based on the the GCDT is constructed as the merging coupling constant zero limit of the stochastic quantizing GCDT model\cite{ALWWZ1}.
Then, it is formulated by a matrix model\cite{ALWWZ2,ALWWZ3}.
In this model the stochastic time plays the role of the geodesic distance\cite{ALWZ2}.
Furthermore, the GCDT model with the additional IK-type interaction is constructed and it is also described by a matrix model formulation\cite{FSW}.
Under the circumstances, in the previous work, we proposed a matrix model formulation of the GCDT with the IK-type interaction based on the loop gas model\cite{Kaw}.
This intuitive analysis leads a new scaling.
Another novelty is that the stochastic time does not correspond to the geodesic distance.

In this paper, we extend the GCDT model for the closed string to the one for the open-closed string, by the extension of the matrix model of the  loop gas model to the matrix-vector model.
In section 2, after reviewing the fundamental nature of the CDT, we construct a matrix-vector model which extend it consistently.
In section 3, we apply it the stochastic quantization method to describe the interactions of loop and line variables.
Though this model contains unsuitable interactions and propagations in the discrete level, in section 4, we find a new scaling in the continuum limit which realizes the open-closed string GCDT model with the IK-type interaction.
The last section is the summary and the conclusion.

\section{CDT matrix-vector model}

In the CDT in 2D space-time, any loop propagator is sliced to many 1-step two-loop functions, each of which is a ring with the small width $a$.
The minimal time, as well as the minimal length, corresponds to the length of the side of the unit triangle $a$.
An 1-step two loop function, or the loop propagator in the unit time, from a loop with $k$ links at the time $t$ to another one with $m$ links at the next time $t+1$, is composed of $k+m$ triangles, $k$ upward triangles and $m$ downward ones.
One site on the loop at $t$ propagates to one or more consecutive links on the loop at $t+1$ and vice versa. 
Assigning a factor $g$ to each triangle and counting the configurations of triangulation, the 1-step propagator is expressed as, 
\begin{eqnarray}
\label{eq:1step}
G^{(0)} (k,m; 1) \equiv {g^{k+m} \over k+m}~_{k+m}{\rm C}_k ,
\end{eqnarray}
where the last factor is the binomial coefficient.
By distinguishing the absolute position of triangles, not only the configuration on the ring, we define another expression, or the 1-step ``marked'' two-loop function,
\begin{eqnarray}
\label{eq:1stepmarked}
 G^{(1)} (k,m;1) \equiv k G^{(0)}(k,m;1) .
\end{eqnarray}
With these 1-step functions, we can construct ``unmarked" and ``marked" two-loop functions of finite $t$-step by the time foliation rule,
\begin{eqnarray}
\label{eq:tstep}
G^{(0)} (n,m;t) & = & \sum_{k=1}^{\infty} G^{(0)}(n,k;t-1) k G^{(0)} (k,m;1), \nonumber \\
G^{(1)} (n,m;t) & = & \sum_{k=1}^{\infty} G^{(1)}(n,k;t-1) G^{(1)} (k,m;1),  
\end{eqnarray}
respectively.
The geodesic distance of the propagation becomes same everywhere on the loop.
It is worth noticing about the time foliation structure that in the CDT we do not have any loop propagation in a same time, or in a ``equi-temporal'' slice.

The causality is violated at the saddle point on the world sheet, where two distinct light-cones are caused.
Although both of splitting and merging interactions should be excluded in the exact sense, we relax this restriction to include only the splitting interaction.
In this regime, branching baby loops eventually shrink to disappear into the vacuum.
In spite of the partial causality violation, the propagating mother loop never interacts with the ill-causality object.
This extended model is the GCDT.

We start with the U($N$) gauge invariant action of a matrix-vector model which is modified from the loop gas model,
\begin{eqnarray}
\label{eq:action}
S[M] & = &  -g\sqrt{N} {\rm tr} \sum_t M_{tt} + {1 \over 2}{\rm tr} \sum_{t,t'} M_{tt'} M_{t't} - {g \over 3\sqrt{N}} {\rm tr} \sum_{t,t',t''} M_{tt'} M_{t't''} M_{t''t} \nonumber \\
 & & + \sum_{t,t'} \sum_{a=1}^R V_t^{a*} \left( {\bf 1}_{tt'} - {g_B^a \over \sqrt{N}} M_{tt'} \right) V_{t'}^a ,
\end{eqnarray}
with the partition function,
$Z = \int {\cal D} M {\cal D} V {\cal D} V^* e^{-S[M,V,V^* ] } $.
We abbreviate indices $i,j$ of an $N \times N$ matrix $( M_{tt'} ) _{ij} ~ (i,j=1 \sim N)$, where the discrete times $t$ and $t'$ are assigned to $i$ and $j$, respectively.
The matrix corresponds to a link directed from a site with $i$ on the time $t$ to another site with $j$ on the time $t'$.
The $N$ dimensional vectors $(V_t^a)_i $ and $(V_t^{a*})_i $ possess one index $i$ attached to the time $t$ and the upper suffix ``$a$'', running from 1 to $R$.
They correspond to the edges of an open line located on the D-brane of ``$a$", in the slice of the time $t$.
We define for $t'=t$, $M_{tt} \equiv A_t$ is an hermitian matrix corresponding to a link in a equi-temporal slice of the time $t$.
For $t'=t \pm 1$, $M_{t,t+1} \equiv B_t $ and $M_{t+1,t} \equiv B^{\dagger}_t $ are the link directed from $t$ to $t+1$ and the one directed from $t+1$ to $t$, respectively.
Otherwise, $M_{tt'}=0$, so that every link connects two sites on same or neighboring times each other.
The action is rewritten with the matrices $A_t , B_t, B_t^{\dagger}$ as
\begin{eqnarray}
\label{eq:actionAB}
S[A, B, B^{\dagger}] &=& -g\sqrt{N} {\rm tr} \sum_t A_{t} + {1 \over 2}{\rm tr} \sum_t A_t^2 + {\rm tr} \sum_t B_t B_t^{\dagger} \nonumber \\
 & & - {g \over 3\sqrt{N}} {\rm tr} \sum_t A_t^3 - {g \over \sqrt{N}} {\rm tr} \sum_t  \left( A_t B_t B_t^{\dagger} + A_{t+1} B_t^{\dagger} B_t  \right) \nonumber \\
 & & + \sum_t \sum_{a=1}^R \left\{  V_t^{a*} V_t^a - {g_B^a \over \sqrt{N}} \left( V_t^{a*} A_t V_t^a + V_t^{a*} B_t V_{t+1}^a + V_{t+1}^{a*} B_t^{\dagger} V_t^a \right) \right\}.
\end{eqnarray}
Let us see the matrix cubic terms in the second line, which correspond to the triangles.
The last two terms composed of $A_t$, $B_t$ and $B_t^{\dagger}$ are elements of the ring of 1-step two-loop function, whereas the first term, cubic only of $A_t$, corresponds to a triangle soaked in one time slice.
It causes the loop propagation in a equi-temporal slice, which is not included in the GCDT.
The quadratic terms in the first line glue the sides of triangles.
While the trace of $B_t B_t^{\dagger}$ connects two triangles to compose a ring of the 1-step two-loop function, the trace of $A_t^2$ connects two links of neighboring rings, by the integration in 
$Z = \int {\cal D} A {\cal D} B {\cal D} B^{\dagger} {\cal D} V {\cal D} V^* e^{-S[A,B,B^{\dagger},V,V^* ] }$.
After integrating out the matrices $B$ and $B^{\dagger}$, we obtain the effective action,
\begin{eqnarray}
\label{eq:effAV}
S_{\rm eff} [A,V,V^*] & = & {\rm tr} \sum_t \left[ -g \sqrt{N} A_t + {1 \over 2} A_t^2  - {g \over 3\sqrt{N}} A_t ^3 
 + \sum_a V_t^{a*} \left( {\bf 1} - {g_B^a \over \sqrt{N}}A_t \right) V_t^a \right. \nonumber \\ 
 & & + \log \Bigl\{ {\bf 1} - {g \over \sqrt{N}}  \left( A_t {\bf 1}_{t+1} + {\bf 1}_t A_{t+1}  \right) \Bigr\} \nonumber \\
 & & \left. - \sum_{a,b} {g_B^a g_B^b \over N} ( V_t^{a*} V_{t+1}^a ) \Bigl\{ {\bf 1} - {g \over \sqrt{N}}  \left( A_t {\bf 1}_{t+1} + {\bf 1}_t A_{t+1}  \right)  \Bigr\}^{-1} ( V_t^b V_{t+1}^{b*} )
\right],
\end{eqnarray}
for the partition function,
$Z = \int {\cal D} A {\cal D} V {\cal D} V^* e^{-S_{\rm eff} [A,V,V^*] } $.
We define the closed loop variable of the length $n$ in the time $t$ as 
$\phi _t (n) \equiv {1 \over N}{\rm tr}({A_t \over \sqrt{N}})^n$
and the open line variable of the length $n$ with the edge factors ``$a$'',``$b$'' in the time $t$ as
$\psi _t^{ab} (n) \equiv {\sqrt{g_B^a g_B^b} \over N} V_t^{a*} \left( {A_t \over \sqrt{N}} \right)^n V_t^b $.
Then, we expand the above effective action (\ref{eq:effAV}) and rewrite it with these variables as
\begin{eqnarray}
\label{eq:effective}
S_{\rm eff} [\phi , \psi , A, V, V^*] & = & S_0 [\phi , \psi , A, V, V^*] + S_1 [ A, V, V^*], \\
\label{eq:S0}
S_0 [\phi , \psi , A, V, V^*] & = & \sum_t \left[ {1 \over 2}{\rm tr}  A_t^2 
- N^2 \sum_{k=0}^{\infty} \sum_{m=0}^{\infty} G^{(0)}(k,m;1) \phi _t (k) \phi _{t+1}(m)  \right.  \nonumber \\
 & & \left. + \sum_a V_t^{a*} V_t^a  - N \sum_{a,b} \sum_{k=0}^{\infty} \sum_{m=0}^{\infty} F^{(0)} (k,m;1) \psi _t ^{ab} (k) \psi _{t+1}^{ba} (m) \right],  \\
\label{eq:S1}
S_1 [A, V, V^*] & = & -N^2 \sum_t \left[ {g \over N} {\rm tr} {A_t \over {\sqrt{N}}} + {g \over 3N}{\rm tr} \left({A_t \over \sqrt{N}}\right)^3 
+ {g_B^a \over N} \sum_a V_t^{a*} {A_t \over \sqrt{N}} V_t^a \right] ,
\end{eqnarray}
where the coefficient of the open line variable quadratic term, 
\begin{eqnarray}
\label{eq:1stepopen}
F^{(0)}(k,m;1) \equiv (k+m)G^{(0)}(k,m;1) = g^{k+m}~_{k+m}{\rm C}_k ,
\end{eqnarray}
is interpreted as the amplitude of 1-step propagation of open line, as the coefficient $G^{(0)}(k,m;1)$ of the closed loop variable quadratic term has the meaning of the 1-step two-loop function.
The finite time $t$-step propagator of the closed loop, or the time foliation of eq.(\ref{eq:tstep}), is expressed as
\begin{eqnarray}
\label{eq:closedpropagator}
nmG^{(0)}(n,m;t) & = & \langle \phi_0 (n) \phi _t (m) \rangle \nonumber \\
 & = & {1 \over Z_0} \int {\cal D} A {\cal D} V {\cal D} V^* \phi _0 (n) \phi _t (m) e^{-S_0 [\phi ,\psi ,A,V,V^*]},
\end{eqnarray}
where the partition function $Z_0 =\int {\cal D}A {\cal D}V {\cal D}V^* e^{-S_0}$ is described with the ``free part'' (\ref{eq:S0}) of the effective action (\ref{eq:effective}).
In the simmilar way, we deduce the $t$-step propagator of the open line as
\begin{eqnarray}
\label{eq:openpropagator}
F^{(0)}(n,m;t)\delta _{ad} \delta _{bc} & = & \langle \psi _0^{ab}(n) \psi _t^{cd}(m) \rangle \nonumber \\
 & = & {1 \over Z_0} \int {\cal D} A {\cal D} V {\cal D} V^* \psi _0^{ab} (n) \psi _t^{cd} (m) e^{-S_0 [\phi ,\psi ,A,V,V^*]},
\end{eqnarray}
Although the extra terms of $S_1$ in the effective action seem to break the time foliation structure at the first sight, they are found to be rather necessary to realize the GCDT structure consistently in the continuum limit.

\section{Stochastic quantization}
We apply the stochastic quantization method to the above model to obtain the GCDT model for open-closed string field theory.
The Langevin equations are
\begin{eqnarray}
\label{eq:langevin}
\Delta (A_t)_{ij} & = & - {{\partial S_{\rm eff}} \over {\partial (A_t)_{ji}}} \Delta \tau + (\Delta \xi _t)_{ij}, \nonumber \\
\Delta (V_t^a)_i & = & - \lambda _t^a {{\partial S_{\rm eff}} \over {\partial (V_t^{a*})_i}} \Delta \tau + (\Delta \eta _t^a)_i, \nonumber \\
\Delta (V_t^{a*})_i & = & - \lambda _t^a {{\partial S_{\rm eff}} \over {\partial (V_t^a)_i}} \Delta \tau + (\Delta \eta _t^{a*})_i, 
\end{eqnarray}
where $\lambda _t^a$ is the scale parameter of the stochastic time evolution on the boundary ``$a$''.
White noise terms $\Delta \xi _t , \Delta \eta _t^a , \Delta \eta _t^{a*}$ satisfy the following correlations:
\begin{eqnarray}
\label{eq:correlation}
\langle (\Delta \xi _t)_{ij} (\Delta \xi _{t'})_{kl} \rangle _{\xi} & = & 2\Delta \tau \delta _{tt'} \delta _{il} \delta _{jk}, \nonumber \\
\langle (\Delta \eta _t^{a*})_i (\Delta \eta _{t'}^b)_j \rangle _{\eta} & = & 2\lambda _t^a \Delta \tau \delta _{tt'} \delta _{ab} \delta _{ij}.
\end{eqnarray}

The Langevin equation for the closed loop variable is
\begin{eqnarray}
\Delta \phi _t (n) & = & \Delta \tau n \left[ g \phi _t (n-1) - \phi _t (n) + g \phi _t (n+1) \right. \nonumber \\
 & & + \sum_{k =0}^{n-2} \phi _t ( k ) \phi_t (n - k - 2)  \nonumber \\
 & & + \sum_{k=1}^{\infty} \sum_{m=0}^{\infty} \left\{  G^{(1)} (k,m;1) \phi _{t+1} (m) + G^{(2)} (m,k;1) \phi _{t-1} (m) \right\} \phi _t (n+k-2)  \nonumber \\
 & & + {1 \over N} \sum_{c,d} \sum_{k=1}^{\infty} \sum_{m=0}^{\infty} \left\{ F^{(0)} (k,m;1) \psi _{t+1}^{dc} (m) + F^{(0)} (m,k;1) \psi _{t-1}^{dc} (m) \right\} k \psi _t^{cd} (n+k-2) \nonumber \\
 & & \left. + {1 \over N} \sum_c \psi _t^{cc} (n-1) \right]
 + \Delta \zeta _t (n),
\end{eqnarray}
where the last term is a constructive noise variable,
$\Delta \zeta _t (n) \equiv {1 \over N} n {\rm tr} \{ {\Delta \xi _t \over \sqrt{N}} ( {A_t \over \sqrt{N}} ) ^{n-1} \}$.
While $G^{(1)} (k,m;1) = kG^{(0)} (k,m;1) $
is the 1-step marked two-loop function with a mark on the entrance loop,
$G^{(2)} (k,m;1) \equiv mG^{(0)} (k,m;1) $
is the one with a mark on the exit loop.

\begin{figure}[t]
 \begin{minipage}{0.47\hsize}
 \vspace{12mm}
  \begin{center}
   \includegraphics [width=45mm]{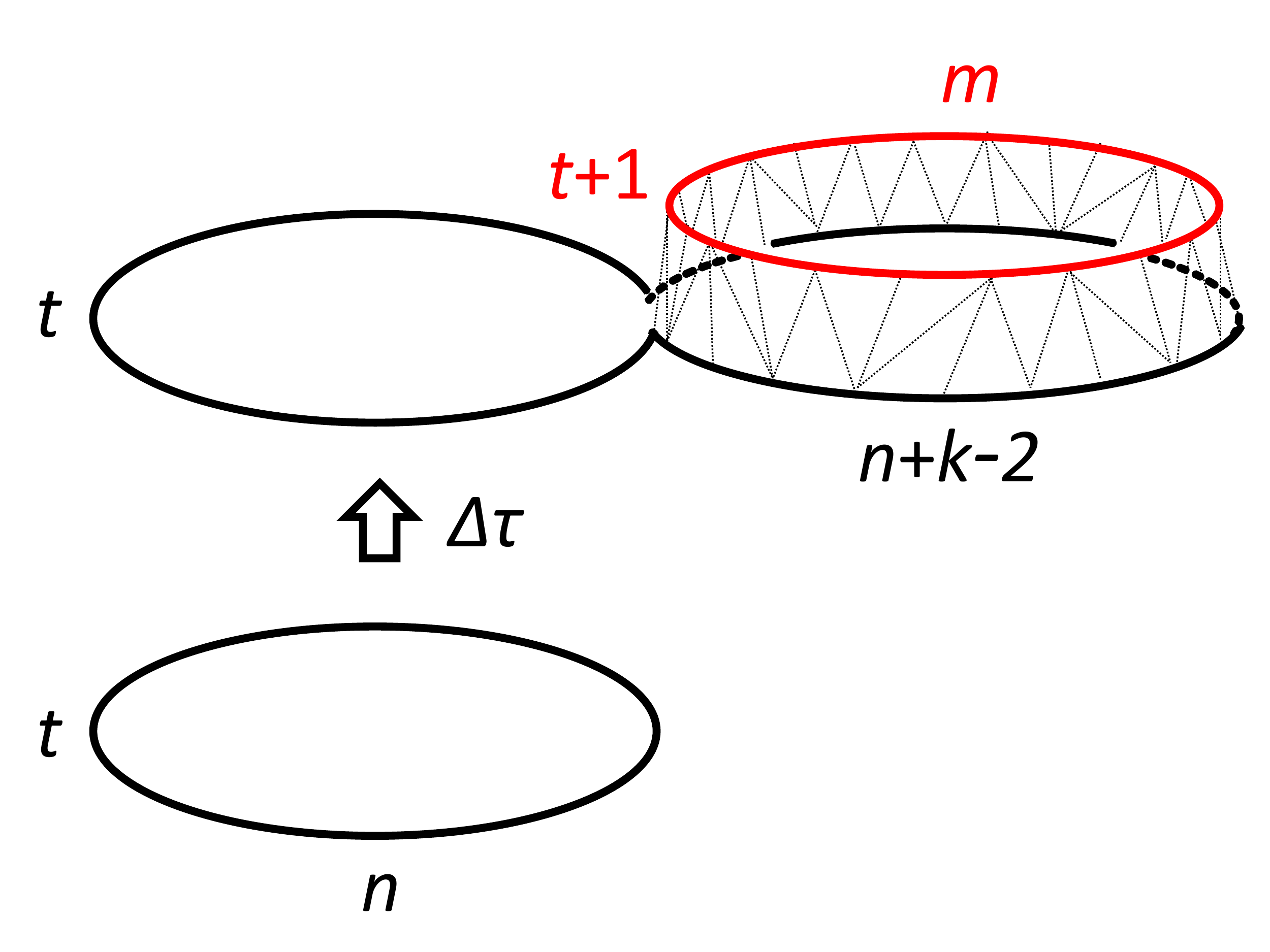}
\end{center}
\caption{IK-type interaction concerning closed loops:
We also have the process of creating a closed loop on the neighboring past time, instead of the future time as above.
}
\label{fig:one}
 \end{minipage}
\hspace{5mm}
 \begin{minipage}{0.47\hsize}
  \begin{center}
   \includegraphics [width=75mm]{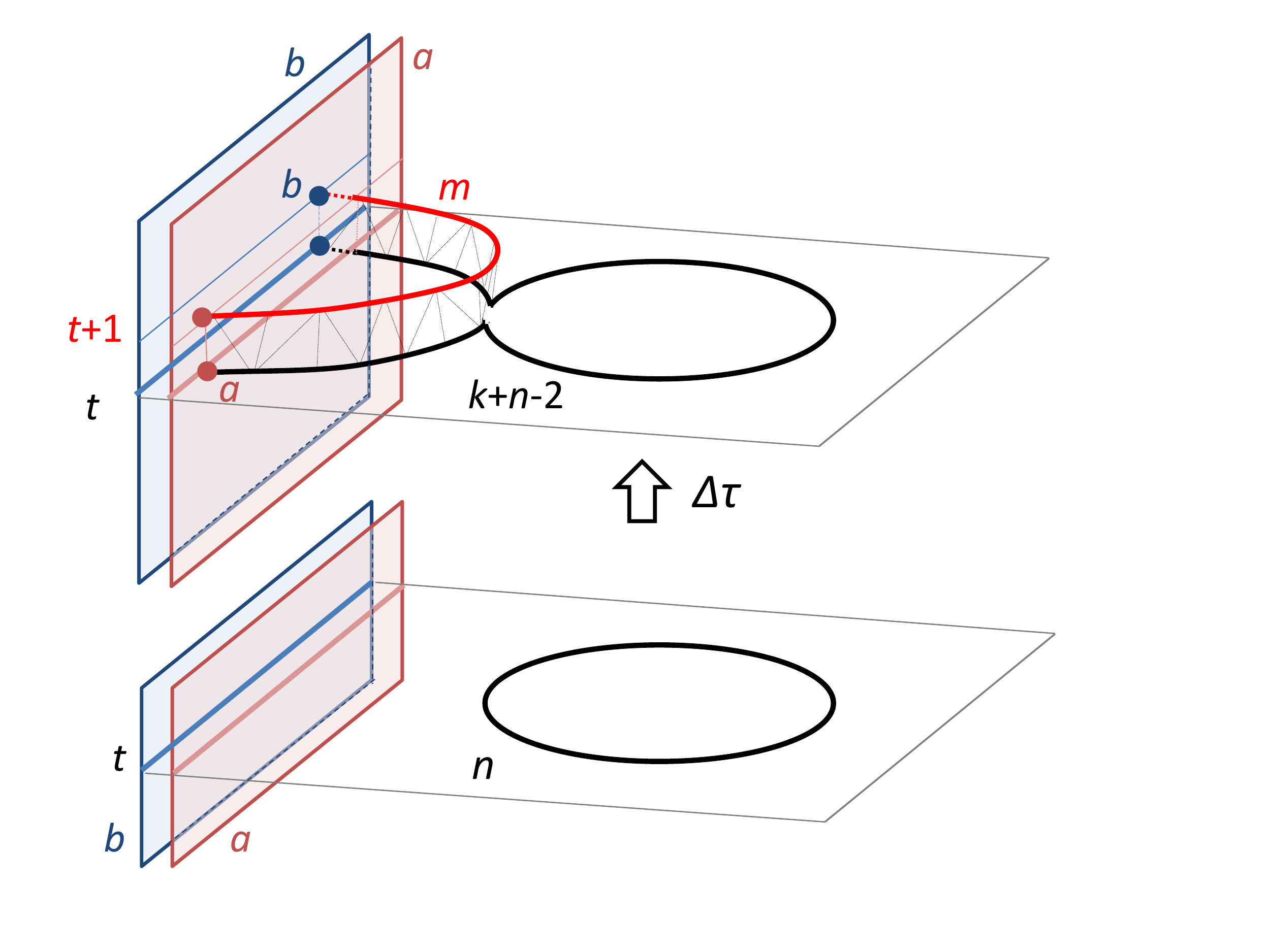}
\end{center}
\caption{IK-type interaction creating open lines from a closed loop}
\label{fig:two}
 \end{minipage}
\end{figure}

The terms in the first line suggest the deformation of the loop in the equi-temporal slice.
The second line is ordinary splitting process.
The third line expresses the IK-type interactions.
These interactions extend the loop length by $k-2$, simultaneously on the neighboring time slice creating a loop with some length $m$, which is related to the extended length $k$ by the 1-step two-loop function(Fig.\ref{fig:one}).
The remains are novel terms including the open line variables.
The first term in the last line means cutting of a closed loop to make an open line.
We interpret the fourth line as the IK-type interactions concerning the pair creation of open lines.
The extensional part, $k$, of the consequential open line and another open line, $m$, created in the neighboring time are related by the 1-step open-line propagator(Fig.\ref{fig:two}).
For the simple expression, we adopt the following abbreviation for the IK-type interactions:
\begin{eqnarray}
\label{eq:IKfactor}
\hat{\phi}_t(k) \equiv \sum_{m=0}^{\infty} \{  G^{(1)} (k,m;1) \phi _{t+1} (m) + G^{(2)} (m,k;1) \phi _{t-1} (m) \}, \nonumber \\
\hat{\psi}_t^{ab}(k) \equiv \sum_{m=0}^{\infty} \{ F^{(0)} (k,m;1) \psi _{t+1}^{ab} (m) + F^{(0)} (m,k;1) \psi _{t-1}^{ab} (m) \}.
\end{eqnarray}

The Langevin equation for the open line variable is 
\begin{eqnarray}
\Delta \psi _t ^{ab} (n) & = & \Delta \tau \left[ n \left\{ g \psi _t^{ab}(n-1) -\psi _t^{ab}(n) +g \psi _t^{ab}(n+1) \right\} \right. \nonumber \\
 & & + \sum_c \sum_{k=0}^{n-1} \psi _t^{ac}(k) \psi _t^{cb}(n-k-1) + n \sum_{k=1}^{\infty} \hat{\phi}_{t}(k) \psi _t^{ab}(n+k-2) \nonumber \\
 & & + \sum_{c,d} \sum_{k=0}^{n-1} \sum_{\ell =0}^{\infty} \sum_{\ell '=0}^{\infty} \hat{\psi}_{t}^{dc}(k) \psi _t^{ad}(k+\ell ') \psi _t^{cb}(n+\ell -k-1) \nonumber \\
 & & \left. + \sum_{k=1}^{n-1} k \psi _t^{ab}(k-1) \phi _t(n-k-1) \right] \nonumber \\
 & & + 2 \lambda _t^a \Delta \tau \delta ^{ab} \phi _t (n) \nonumber \\
 & & + \lambda _t^a \Delta \tau \left[ - \psi _t^{ab}(n) + g_B^a \psi _t^{ab}(n+1) + \sum_c \sum_{k=0}^{\infty} \hat{\psi}_{t}^{ac}(k) \psi _t^{cb}(n+k) \right] \nonumber \\
 & & + \lambda _t^b \Delta \tau \left[ - \psi _t^{ab}(n) + g_B^b \psi _t^{ab}(n+1) + \sum_c \sum_{k=0}^{\infty} \hat{\psi} _{t}^{cb}(k) \psi _t^{ac}(n+k) \right] \nonumber \\
 & & + \zeta _t^{ab}(n),
\end{eqnarray}
where the last term is another constructive noise variable,
\[
\Delta \zeta _t^{ab} (n)  \equiv  {\sqrt{g_B^a g_B^b} \over N} \left\{ \sum_{k=0}^{n-1} V_t^{a*} \! \left({A_t \over \sqrt{N}} \right)^k \! {{\Delta \xi _t} \over \sqrt{N}} \! \left({A_t \over \sqrt{N}} \right)^{n-k-1}  \hspace{-5mm} V_t^b 
 + \! \Delta \eta _t^{a*} \! \left( {A_t \over \sqrt{N}} \right)^n \! V_t^b + \! V_t^{a*} \! \left( {A_t \over \sqrt{N}} \right)^n \! \Delta \eta _t^b \right\} \! .
\]
The above constructive noise variables satisfy the following correlations:
\begin{eqnarray}
\langle \Delta \zeta _t (n) \Delta \zeta _{t'} (m) \rangle _{\xi} & = & 2 \Delta \tau \delta _{tt'} {1 \over N^2} nm \langle \phi _t (n+m-2) \rangle _\xi , \nonumber \\
\langle \Delta \zeta _t^{ab}(n) \Delta \zeta _{t'}^{cd}(m) \rangle _{\xi \eta} & = & 2\Delta \tau \delta _{tt'}{1 \over N}
\left\{ \lambda ^a_t \delta ^{ad} \langle \psi _t^{cb}(n+m) \rangle _{\eta} + \lambda ^b_t \delta ^{cb} \langle \psi _t^{ad}(n+m) \rangle _{\eta} \right. \nonumber \\
 & & \left. + \sum_{k=0}^{n-1} \sum_{\ell =0}^{m-1} \langle \psi _t^{ad}(k+\ell ) \rangle _{\xi} \langle \psi _t^{cb}(n+m-k-\ell -2) \rangle _{\xi} \right\} \nonumber \\
\langle \Delta \zeta _t (n) \Delta \zeta _{t'}^{ab} (m) \rangle _{\xi} & = & 2 \Delta \tau \delta _{tt'} nm \langle \psi _t^{ab} (n+m-2) \rangle _{\xi} .
\end{eqnarray}
These noise correlations provide us with the merging and cross-changing processes, which should be avoided by the causality, in the stochastic time evolution.
Here, we consider some observable $O(\phi , \psi)$ composed of loop and line variables.
When $O(\phi(\tau +\Delta \tau) , \psi (\tau +\Delta \tau ))$ is expanded around $\tau$, the Fokker-Planck (FP) Hamiltonian is defined as the generator for the stochastic time evolution of the expectation value,
\begin{eqnarray}
\label{eq:FPH}
\langle \Delta O(\phi, \psi ) \rangle _{\xi \eta}  \equiv  - \Delta \tau \langle H_{\rm FP} O(\phi , \psi ) \rangle _{\xi \eta} + O( \Delta \tau ^{3 \over 2}).
\end{eqnarray}
We interpret $\phi _t (n)$ (and $\psi _t^{ab}(n)$) as the creation operators of closed loop (and open line with edges on ``$a$'' and ``$b$'') with the length $n$ in the time $t$, while $\pi _t (n) \equiv {\partial \over \partial \phi _t (n) }$ (and $\pi _t^{ab} (n) \equiv {\partial \over \partial \psi _t^{ab} (n) }$) as the annihilation operators of corresponding loop (and line).
Of course, they satisfy the following commutation relations:
\begin{eqnarray}
\label{eq:commutation}
[\pi _t (n), \phi _{t'} (m) ] = \delta _{tt'} \delta _{nm} ,~~~~
[\pi _t^{ab} (n), \psi _{t'}^{cd} (m) ] = \delta _{tt'} \delta _{ac} \delta _{bd} \delta _{nm}.
\end{eqnarray}
The FP Hamiltonian is expressed in the form,
\begin{eqnarray}
\label{eq:DFPH}
H_{\rm FP} & = & \sum_t \left[ {1 \over N^2} \sum_{n=1}^{\infty} n L_t(n-2) \pi _t (n) \right. \nonumber \\
 & & + {1 \over N} \sum_{ab} \sum_{n=0}^{\infty} \left\{ \lambda ^a_t J_t^{ab}(n) + \lambda^b_t J_t^{ba*}(n) \right\} \pi _t^{ab} (n) \nonumber \\
 & & +  \sum_{ab} \sum_{n=1}^{\infty} \left\{ {1 \over N} \sum_c \sum_{k=1}^{n-1} k \left( J_t^{cb}(k-1) \psi _t^{ac}(n-k-1) + J_t^{ca*}(n-1) \psi _t^{cb}(n-k-1) \right) \right. \nonumber \\
 & & \left. \left. + n K_t^{ab}(n-2) \right\} \pi _t^{ab}(n) \right],
\end{eqnarray}
with three generators $L_t(n)$, $J_t^{ab}(n)$ and $K_t^{ab}(n)$.
The first line with the generator $L_t(n)$ contains the stochastic processes of the closed loop $\phi _t (n)$.
The second line with the generator $J^{ab}_t(n)$ corresponds to the deformation on the edges of the open line $\psi _t^{ab}(n)$.
\footnote{While the generator $J_t^{ab}(n)$ concerns the processes on the edge of ``$a$'' side, $J_t^{ba*}(n)$ is the one at ``$b$'' side.
Notice that the hermitian matrix model constructs the orientable string model.
The complex conjugate means the reversal of the orientation of the link.}
The third line and the fourth line including the generator $K_t^{ab}(n)$ are the processes occurring at some point except at the edges, of the same open line.
In the discrete level, the three generators express the algebraic structure including the Virasoro algebra and SU($R$) current algebra, associated with the model of string with $R$ D-branes located at the same position.
We will see the explicit form of three generators and their commutators in the appendix.

\section{Continuum limit}

In the discrete model, we obtain not only the GCDT processes but also the extra ones inappropriate from the criteria of the causality and the time foliation structure.
We expect these ill-processes to scale out in the continuum limit.
In the double scaling limit, the minimum scale $a$ of length and time goes to zero as $N$ grows to infinity.
According to the CDT structure, the finite length $L$ and finite time $T$ scale in the same way as
\begin{eqnarray}
L \equiv an,~~~~
T \equiv at.
\end{eqnarray}
The infinitesimal expression of the 1-step propagators is
\begin{eqnarray}
\tilde{G}^{(1)}(L, L' ; a) \equiv a^{-1} G^{(1)}(k,m;1) ,~~~~
\tilde{F}^{(0)}(L , L' ; a ) \equiv a^{-1} F^{(0)}(k,m;1) .
\end{eqnarray}
The cosmological constant $\Lambda$ and the boundary cosmological constant $x^a$ are defined from the matrix-vector model coupling constants $g$ and $g_B^a$, respectively, as
\begin{eqnarray}
{1 \over 2}{\rm e}^{-{1 \over 2}a^2 \Lambda} \equiv g,~~~~{\rm e}^{-a x^a } \equiv g_B^a. 
\end{eqnarray}
Based on the above scaling, we define two parameters $D$ and $D_N$, or scaling dimensions, and investigate the range in the parameter space for the realization of the GCDT open-closed string field theory.
The string coupling constant $G_{\rm st}$ is defined with one scaling dimension $D_N$ as
\begin{eqnarray}
 G_{\rm st} = a^{D_N}{1 \over N^2}.
\end{eqnarray}
With another scaling dimension $D$, the definition of the infinitesimal stochastic time $d\tau$ and the boundary scale parameter $\lambda ^a$ is, 
\begin{eqnarray}
\label{eq:stochastictime}
d \tau = a^{{1 \over 2}D-2}\Delta \tau  ,~~~~ \lambda ^a =a^{-{1 \over 4}D+{3 \over 2}}\lambda _t^a .
\end{eqnarray}
We redefine the creation operator $\Phi (L;T)$ and the annihilation operator $\Pi (L;T)$ for the closed string as
\begin{eqnarray}
\Phi (L;T) = a^{-{1 \over 2}D} \phi _t (n) ,~~~~ \Pi (L;T) = a^{{1 \over 2}D-2} \pi _t (n) , 
\end{eqnarray}
in addition, the creation operator $\Psi ^{ab}(L;T)$ and the annihilation operator $\Pi ^{ab}(L;T)$ for the open string as
\begin{eqnarray}
\Psi ^{ab}(L;T) = a^{-{1 \over4}D-{1 \over 2}} \psi _t^{ab}(n) ,~~~~ \Pi ^{ab}(L;T) = a^{{1 \over 4}D-{3 \over 2}} \pi _t^{ab}(n) ,
\end{eqnarray}
in accordance with the commutation relations,
\begin{eqnarray}
\left[ \Pi (L;T), \Phi (L';T') \right] &=& \delta (T-T') \delta (L-L') , \nonumber \\
\left[ \Pi ^{ab}(L;T) , \Psi ^{cd}(L';T') \right] &=& \delta ^{ac} \delta ^{bd} \delta (T-T') \delta (L-L').
\end{eqnarray}
The scaling of the abbreviated form concerning the IK-type interaction, $\hat{\Phi}(L;T)$ and $\hat{\Psi}^{ab}(L;T)$, is also defined consistently,
\begin{eqnarray*}
\hat{\Phi}(L';T) &\equiv& \int_0^{\infty} dL'' \tilde{G}^{(1)}(L',L'';a) \Phi (L'';T+a) + \int_0^{\infty} dL'' \tilde{G}^{(2)}(L'',L';a) \Phi (L'';T-a), \\
\hat{\Psi}^{ab}(L';T) &\equiv& {\int_0^{\infty}  dL'' \tilde{F}^{(0)}(L',L'';a) \Psi^{ab} (L'';T+a)} + \int_0^{\infty}  dL'' \tilde{F}^{(0)}(L'',L';a) \Psi^{ab} (L'';T-a) .
\end{eqnarray*}
At this point, in order for the  minimal stochastic time to become infinitesimal, from eq.(\ref{eq:stochastictime}), the parameter $D$ is restricted to $D>4$.
The continuum limit of the FP Hamiltonian ${\cal H}_{\rm FP}$, which is defined by ${\cal H}_{\rm FP} d \tau \equiv H_{\rm FP} \Delta \tau $, is as follows:
\begin{eqnarray}
\label{eq:FPHamiltonian}
{\cal H}_{\rm FP} = {\cal H}_1 +{\cal H}_2 + {\cal H}_{2'} + {\cal H}_3,
\end{eqnarray}
with 
\begin{eqnarray}
{\cal H}_1 & = & \int dT \int_0^{\infty} dL L \left[ a^{-{1 \over 2}D+3} {1 \over 2}\left( {\partial ^2 \over \partial L^2} - \Lambda \right) \Phi (L;T) \right. \label{eq:FPH1} \\
&&+ a^{-{3 \over 4}D+{3 \over 2}-{1 \over 2}D_N} \sqrt{G_{\rm st}} \sum_c \Psi ^{cc}(L;T) \label{eq:FPH2} \\
&& + \int_0^L dL' \Phi (L';T) \Phi (L-L';T) \label{eq:FPH3} \\
&& + a^{-D+1-D_N} G_{\rm st} \sum_{cd} \int_0^{\infty} dL' L' \Phi (L+L';T) \Pi (L';T) \label{eq:FPH4} \\
&& + \int_0^{\infty} dL' \Phi (L+L';T) \hat{\Phi}(L';T) \label{eq:FPH5} \\
&& + a^{-D+1-D_N} G_{st} \sum_{cd}  \int_0^{\infty} dL' L' \Psi ^{cd}(L+L';T) \Pi ^{cd}(L';T) \label{eq:FPH6} \\
&& \left. + a^{-{1 \over 2}D-{1 \over 2}D_N} \sqrt{G_{st}} \sum_{cd} \int_0^{\infty} dL' L' \Psi ^{cd}(L+L';T) \hat{\Psi}^{dc}(L';T) \right] \Pi (L;T) , \label{eq:FPH7} 
\end{eqnarray}
\begin{eqnarray}
{\cal H}_2 & = & \sum_{ab} \lambda ^a \int dT \int_0^{\infty} dL \left[  a^{-{1 \over 4}D+{3 \over 2}} \left({\partial \over \partial L} - x^a \right) \Psi ^{ab}(L;T) \right. \label{eq:FPH8} \\
&& + \delta ^{ab} \Phi (L;T) \label{eq:FPH9} \\
&& + a^{-{1 \over  2}D+1-{1 \over 2}D_N} \sqrt{G_{\rm st}} \sum_c \int_0^{\infty} dL' \Psi ^{cb}(L+L';T) \Pi ^{ca}(L';T) \label{eq:FPH10} \\
&& \left. + \sum_c \int_0^{\infty} dL' \Psi ^{cb}(L+L';T) \hat{\Psi}^{ac}(L';T) \right] \Pi^{ab}(L;T) , \label{eq:FPH11} 
\end{eqnarray}
\begin{eqnarray*}
{\cal H}_{2'} & = & \sum_{ab} \lambda ^b \int dT \int_0^{\infty} dL \left[ a^{-{1 \over 4}D+{3 \over 2}} \left({\partial \over \partial L} - x^b \right) \Psi ^{ab}(L;T) \right. \\
&&  + \delta ^{ab} \Phi (L;T)  \\
&& + a^{-{1 \over  2}D+1-{1 \over 2}D_N} \sqrt{G_{\rm st}} \sum_c \int_0^{\infty} dL' \Psi ^{ac}(L+L';T) \Pi ^{bc}(L';T)  \\
&& \left. + \sum_c \int_0^{\infty} dL' \Psi ^{ac}(L+L';T) \hat{\Psi}^{cb}(L';T)  \right] \Pi^{ab}(L;T), 
\end{eqnarray*}
\begin{eqnarray}
{\cal H}_3 & = & \sum_{ab} \int dT \int_0^{\infty} dL \left[ a^{-{1 \over 2}D+3} {1 \over 2}L \left( {\partial ^2 \over \partial L^2} - \Lambda \right) \Psi ^{ab} (L;T) \right. \label{eq:FPH12} \\
&& + a^{-{1 \over 4}D+{3 \over 2}} \sum_c \int_0^{\infty} dL' \Psi ^{ac}(L';T) \Psi ^{cb}(L-L';T) \label{eq:FPH13} \\
&& + L \int_0^{\infty} dL' \Psi ^{ab}(L+L';T) \hat{\Phi}(L';T) \label{eq:FPH14} \\
&& + \sum_{cd} \int_0^L  dL' \int_0^{\infty} dL'' \int_0^{\infty} dL''' \nonumber \\
&& \hspace{35mm} \Psi ^{ad}(L'+L''';T) \Psi ^{cb}(L+L''-L';T) \hat{\Psi}^{dc}(L''+L''';T) \label{eq:FPH15} \\
&& + 2 \int_0^L dL' L' \Psi ^{ab}(L';T) \Phi (L-L';T) \label{eq:FPH16} \\
&&  + a^{-{1 \over 2}D+1-{1 \over 2}D_N}  \sqrt{G_{st}} \sum_{cd} \int_0^{\infty} dL'  \int_0^{\infty} dL''  \int_0^{L'} dL'''  \nonumber \\
&& \hspace{35mm} \Psi ^{ad}(L''+L''';T) \Psi ^{cb} (L+L'-L''-L''';T) \Pi ^{cd}(L';T)  \label{eq:FPH17} \\
&& \left. + a^{-D+1-D_N} G_{\rm st} L \int_0^{\infty} dL' L' \Psi ^{ab}(L+L';T) \Pi (L';T) \right] \Pi ^{ab} (L;T) . \label{eq:FPH18} 
\end{eqnarray}
\begin{figure}
$\bullet {\cal H}_1:$ Deformation of closed string
\\
 \begin{minipage}{0.24\hsize}
  \begin{center}
   \includegraphics[width=32mm]{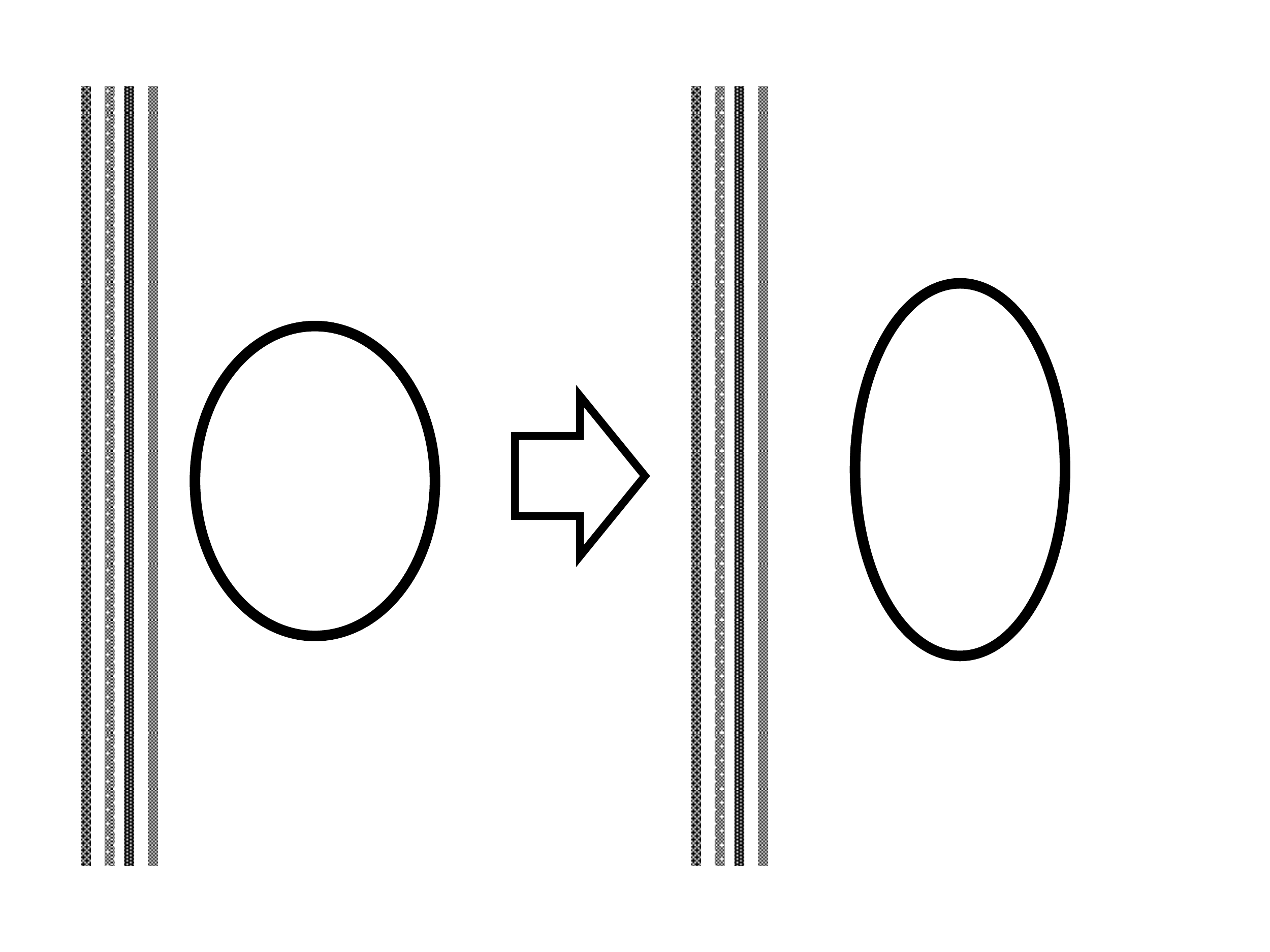}
  \\
(\ref{eq:FPH1})~~Propagation\\
 in a time $\times$
\end{center}

 \end{minipage}
 \begin{minipage}{0.24\hsize}
  \begin{center}
   \includegraphics[width=32mm]{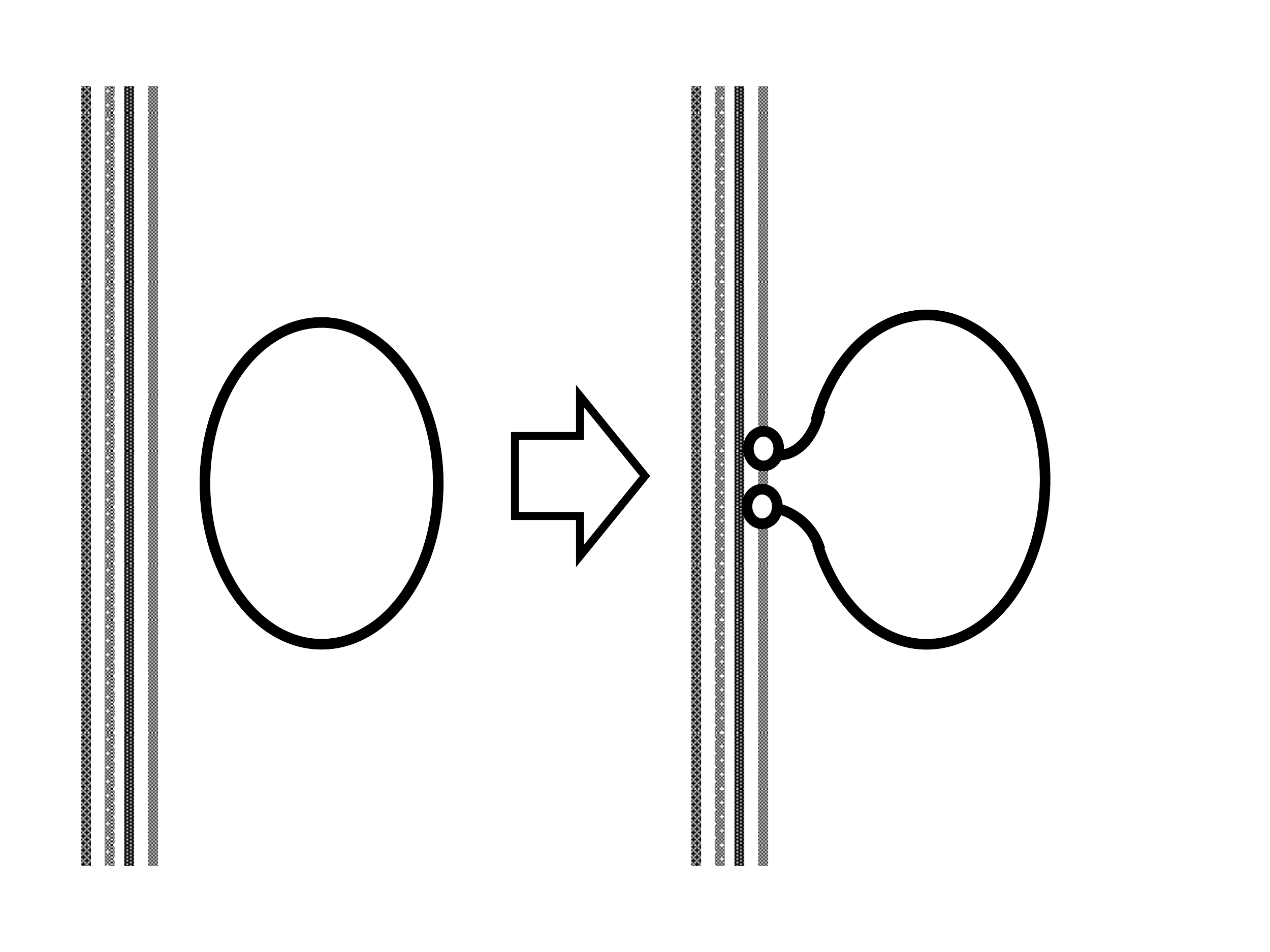}
  \\
(\ref{eq:FPH2})~~Closed string \\
$\rightarrow$ Open string $\times$
\end{center}
 \end{minipage}
 \begin{minipage}{0.24\hsize}
  \begin{center}
   \includegraphics[width=32mm]{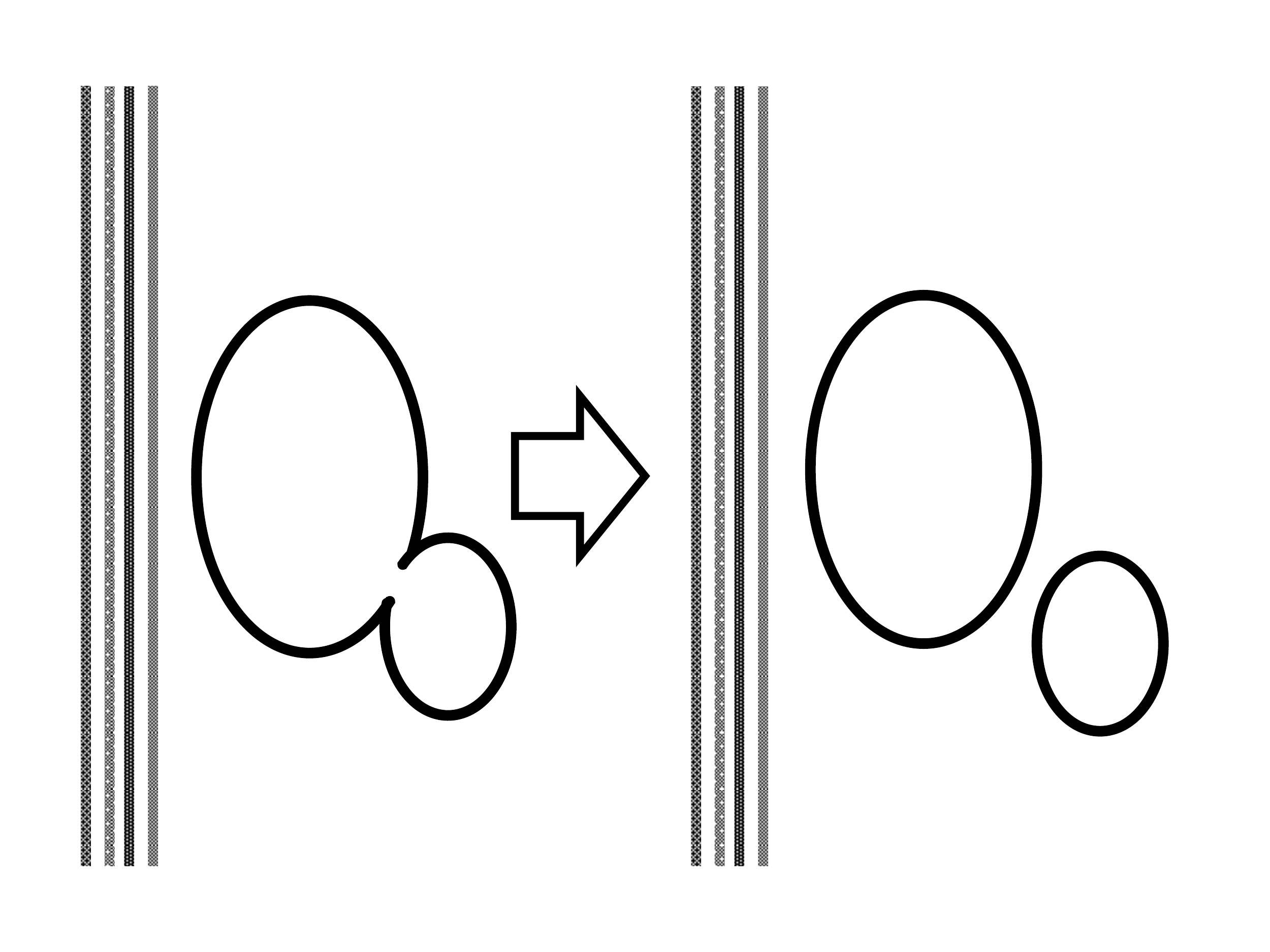}
   \\
(\ref{eq:FPH3})~~Splitting $\bigcirc$
\end{center}
 \end{minipage}
 \begin{minipage}{0.24\hsize}
  \begin{center}
   \includegraphics[width=32mm]{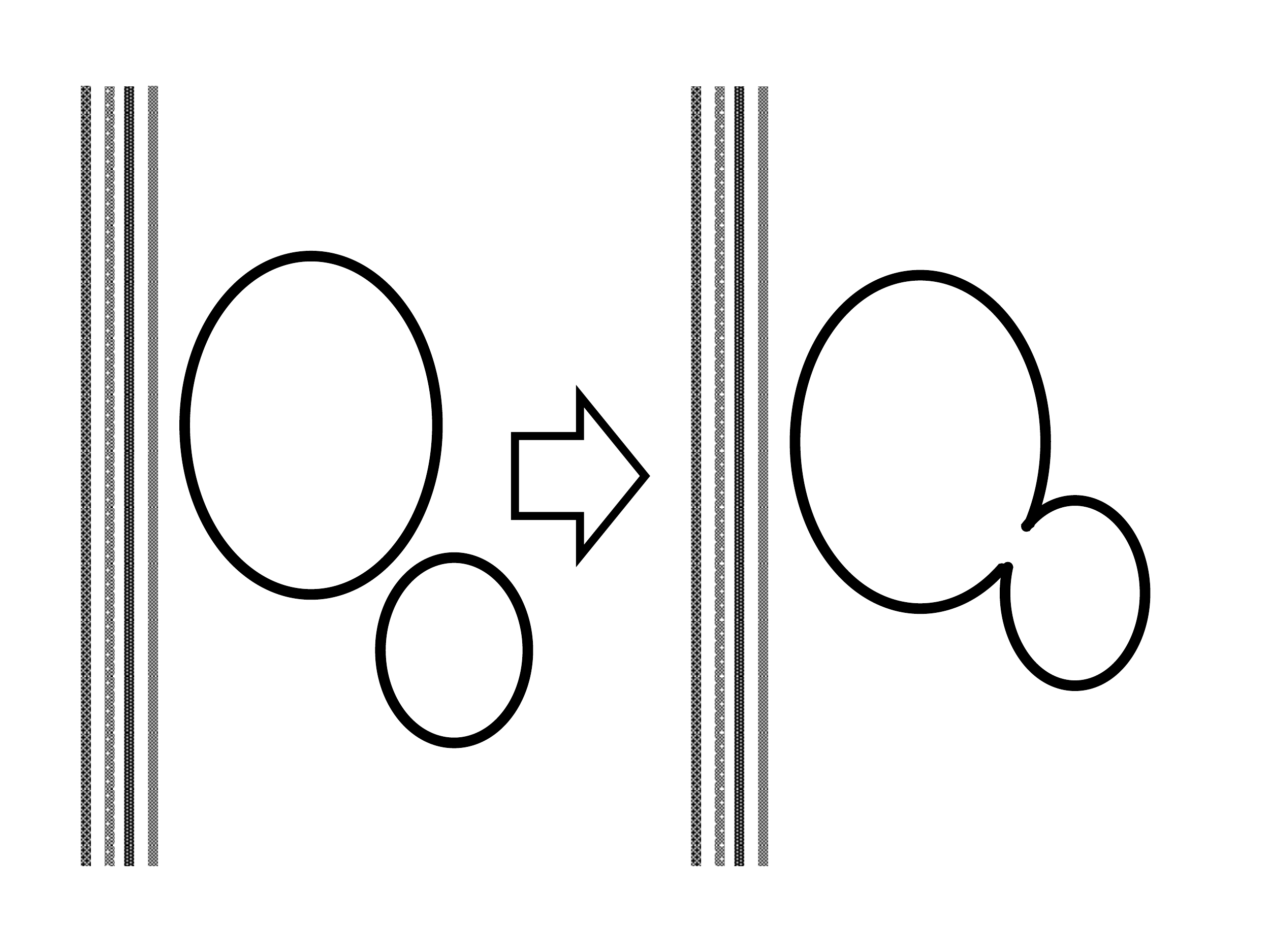}
   \\
(\ref{eq:FPH4})~~Merging $\times$
\end{center}
 \end{minipage}
\\
\begin{minipage}{0.24\hsize}
  \begin{center}
   \includegraphics[width=32mm]{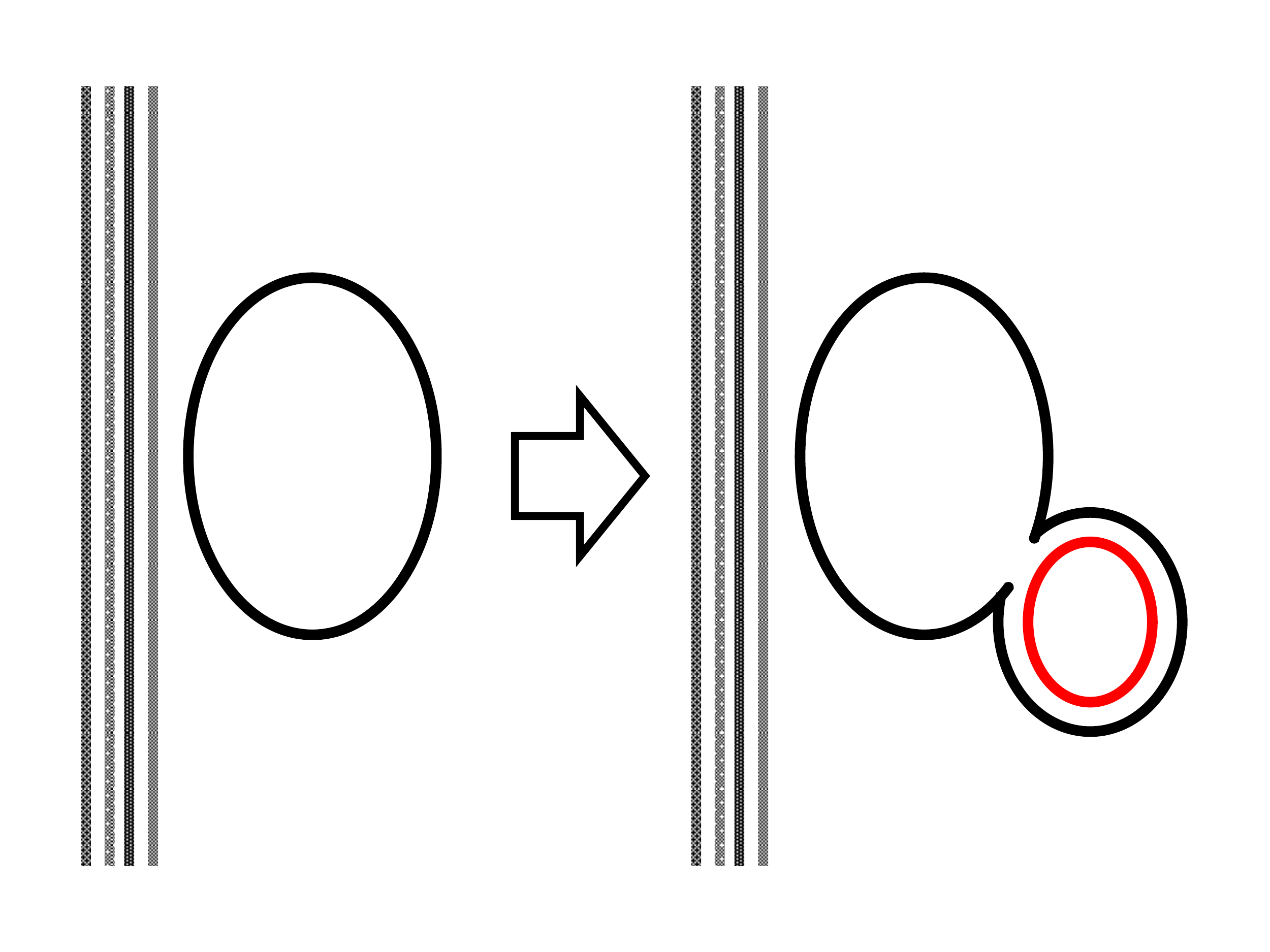}
   \\
(\ref{eq:FPH5})~~IK-type with\\
 closed string $\bigcirc$
\end{center}
 \end{minipage}
 \begin{minipage}{0.24\hsize}
  \begin{center}
   \includegraphics[width=32mm]{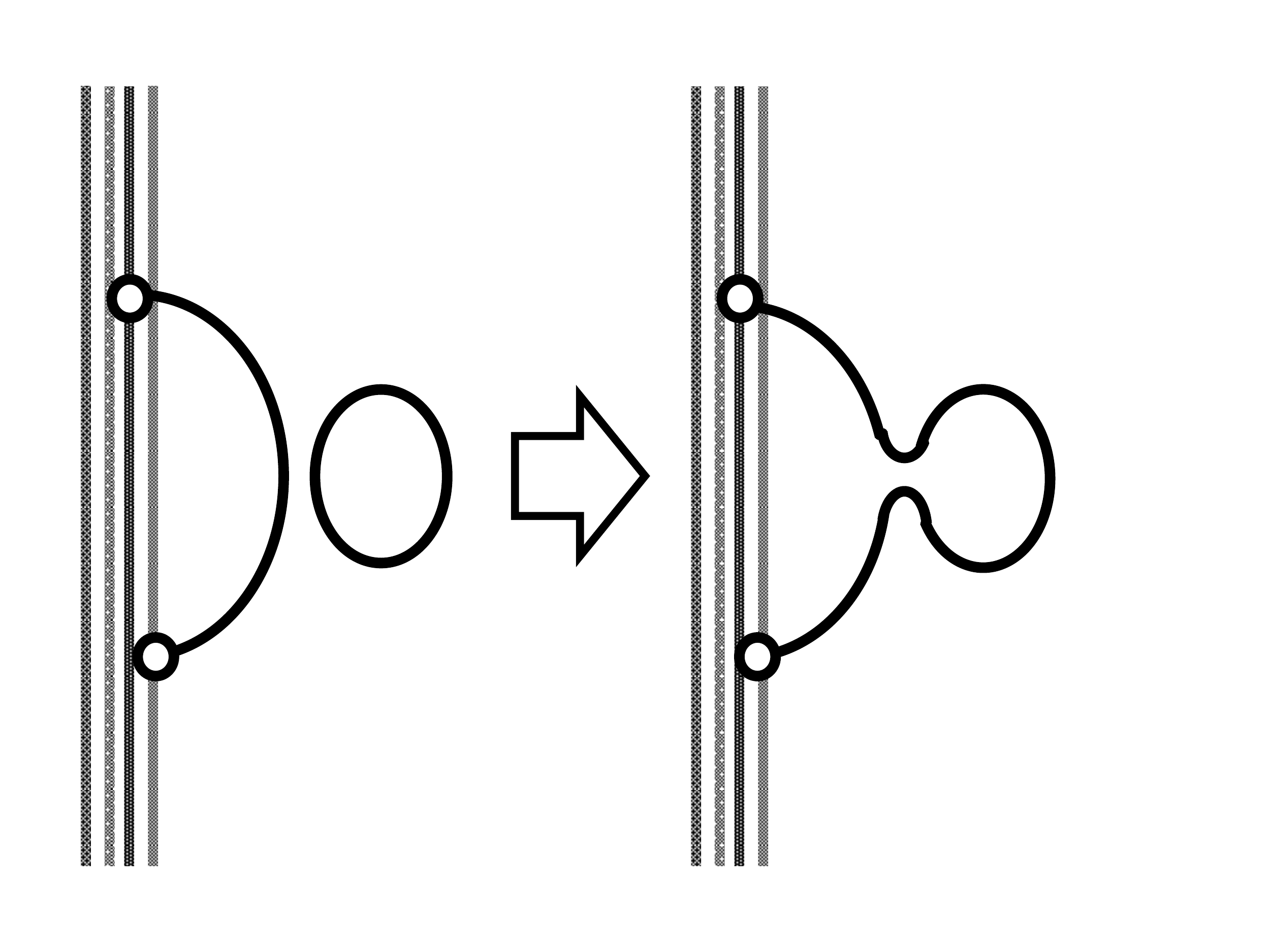}
   \\
(\ref{eq:FPH6})~~Merging with\\
 open string $\times$
\end{center}
 \end{minipage}
 \begin{minipage}{0.24\hsize}
  \begin{center}
   \includegraphics[width=32mm]{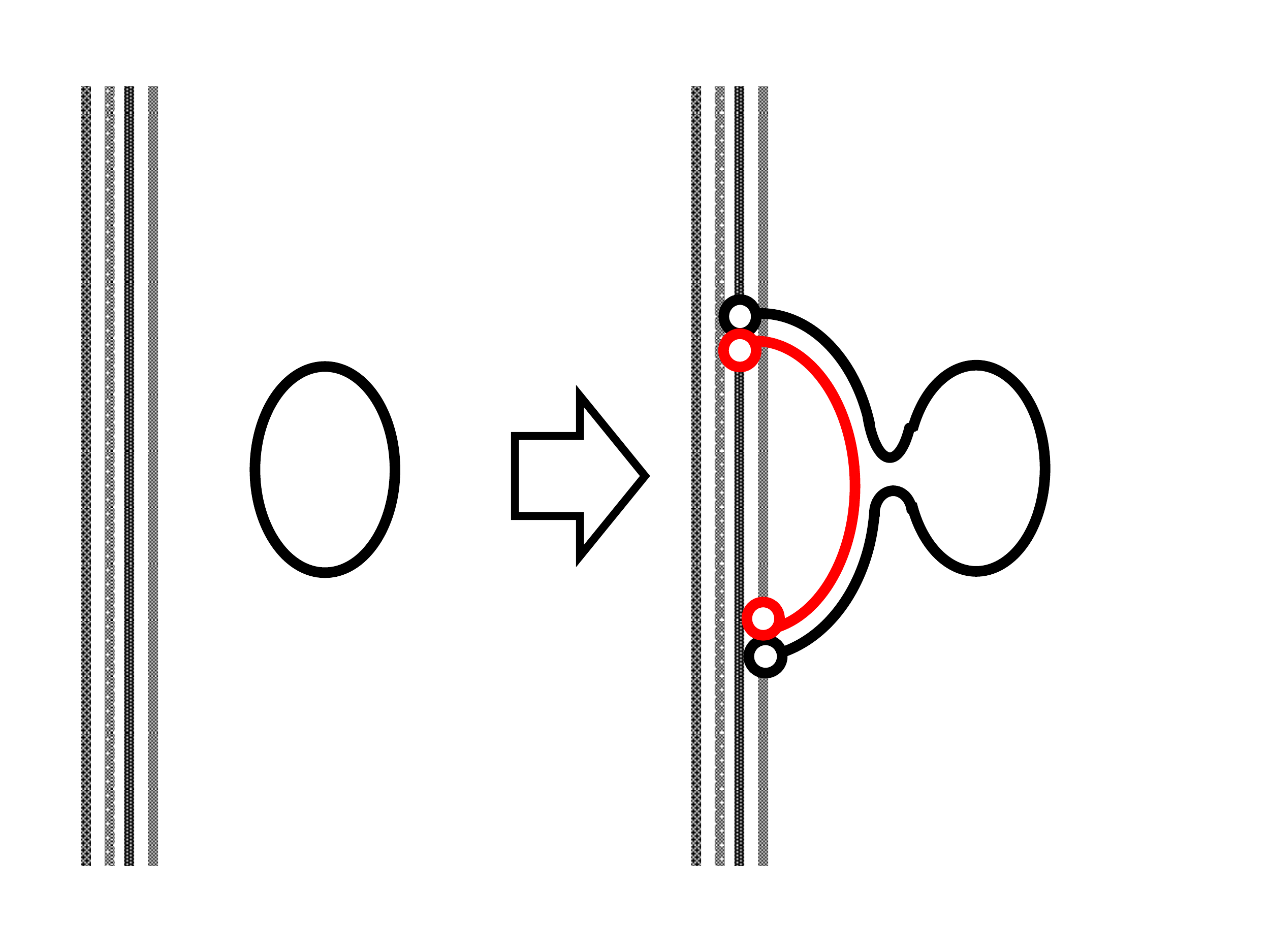}
   \\
(\ref{eq:FPH7})~~IK-type with\\
 open string $\bigtriangleup$
\end{center}
 \end{minipage}
 \begin{minipage}{0.24\hsize}
 \end{minipage}
\\
\vspace{5mm}
\\
$\bullet {\cal H}_2:$ Deformation at the edge of open string
\\
 \begin{minipage}{0.24\hsize}
  \begin{center}
   \includegraphics[width=32mm]{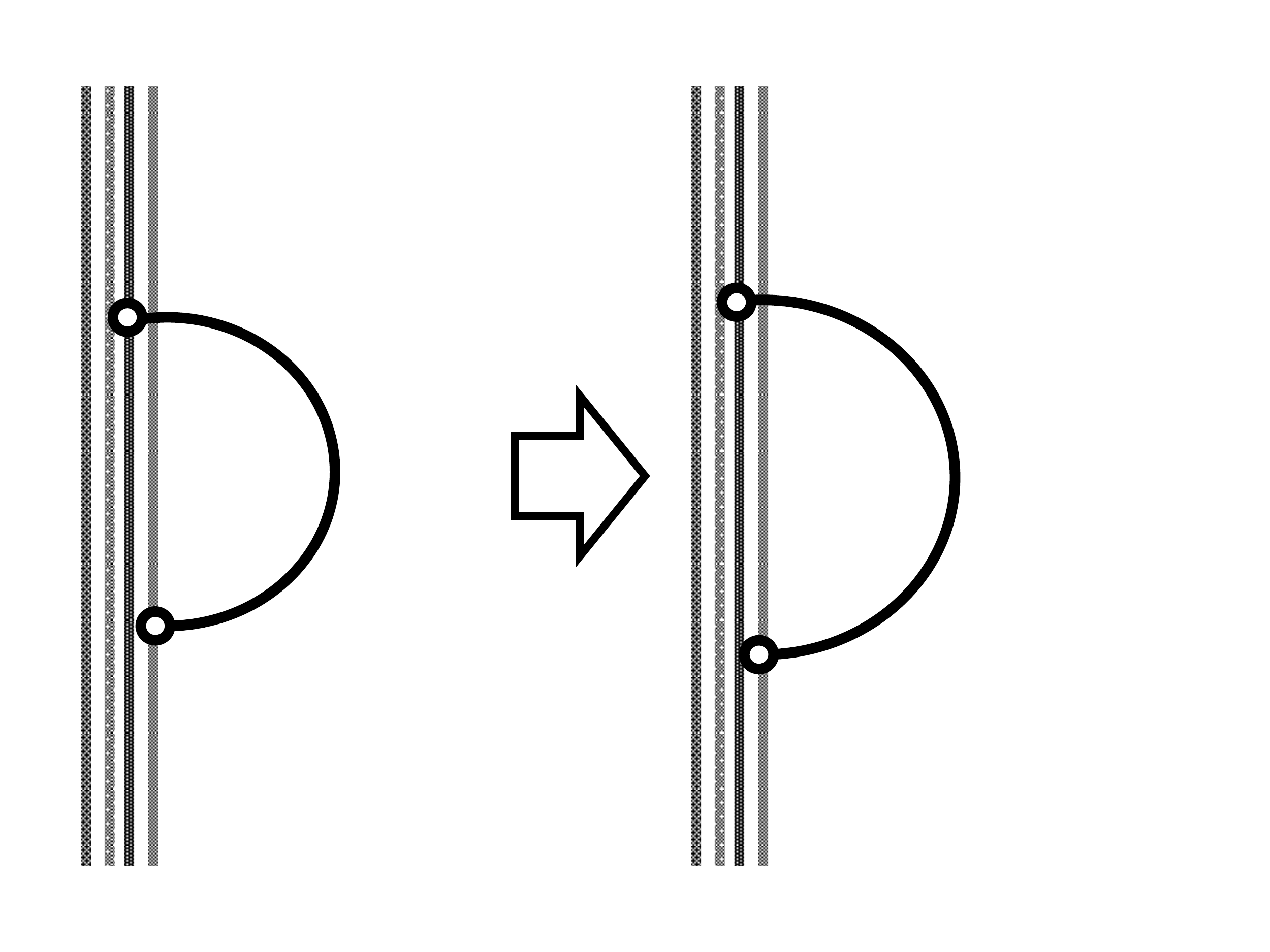}
   \\
(\ref{eq:FPH8})~~Propagation\\
 in a time $\times$
\end{center}
 \end{minipage}
 \begin{minipage}{0.24\hsize}
  \begin{center}
   \includegraphics[width=32mm]{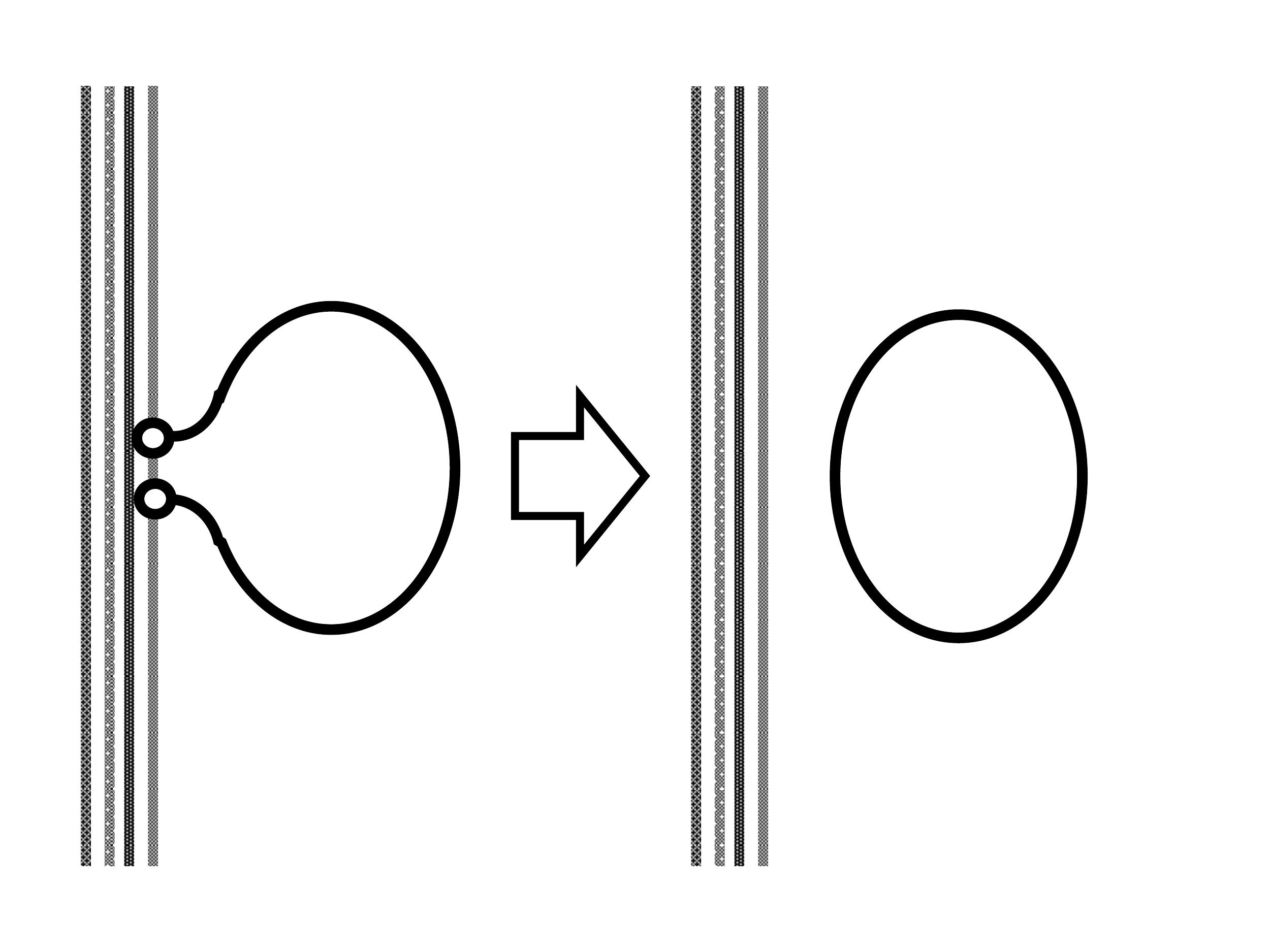}
   \\
(\ref{eq:FPH9})~~Open string $\rightarrow$ \\
Closed string $\bigcirc$
\end{center}
 \end{minipage}
 \begin{minipage}{0.24\hsize}
  \begin{center}
   \includegraphics[width=32mm]{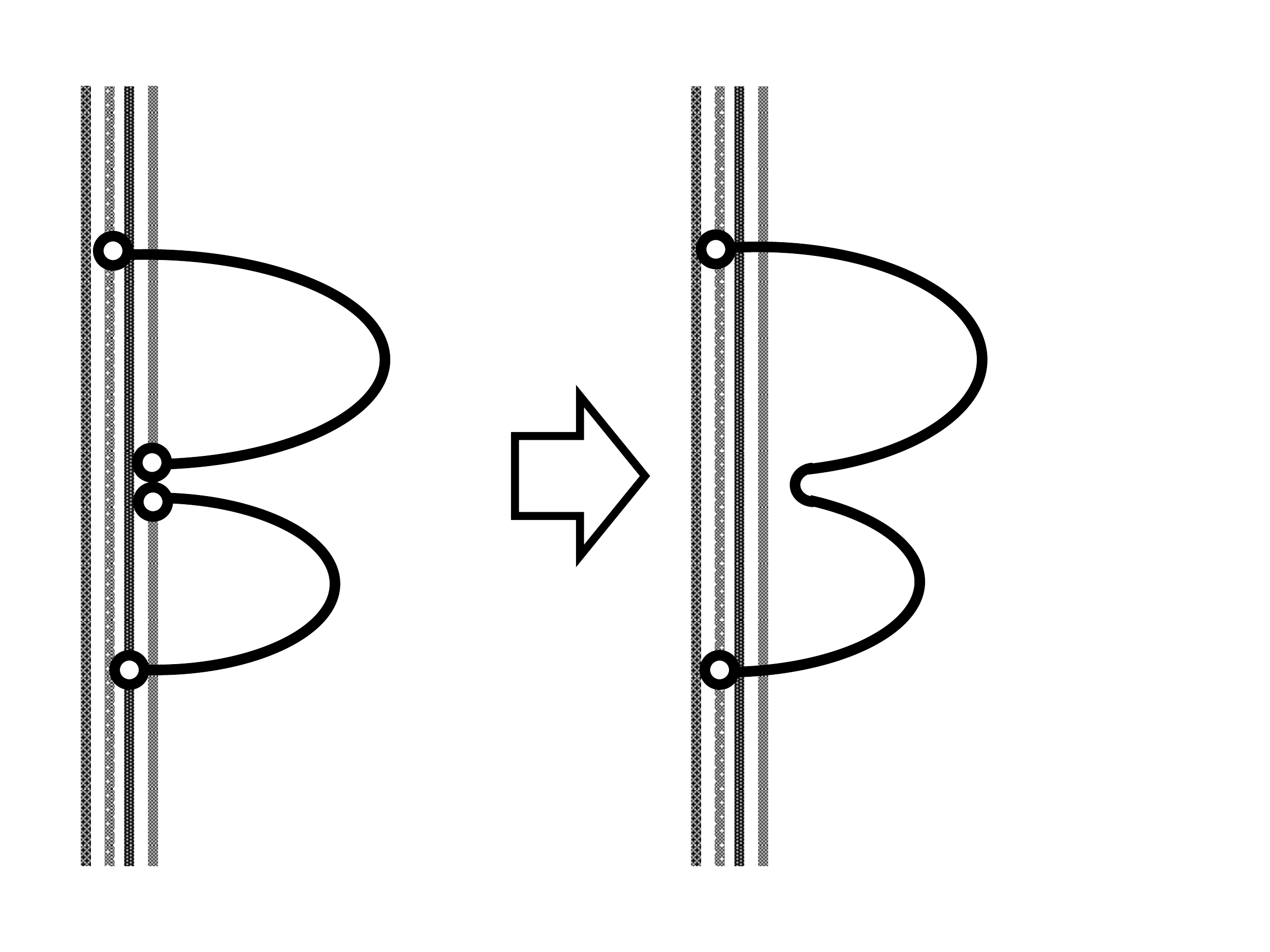}
   \\
(\ref{eq:FPH10})~~Merging with\\
 open string $\times$
\end{center}
 \end{minipage}
 \begin{minipage}{0.24\hsize}
  \begin{center}
   \includegraphics[width=32mm]{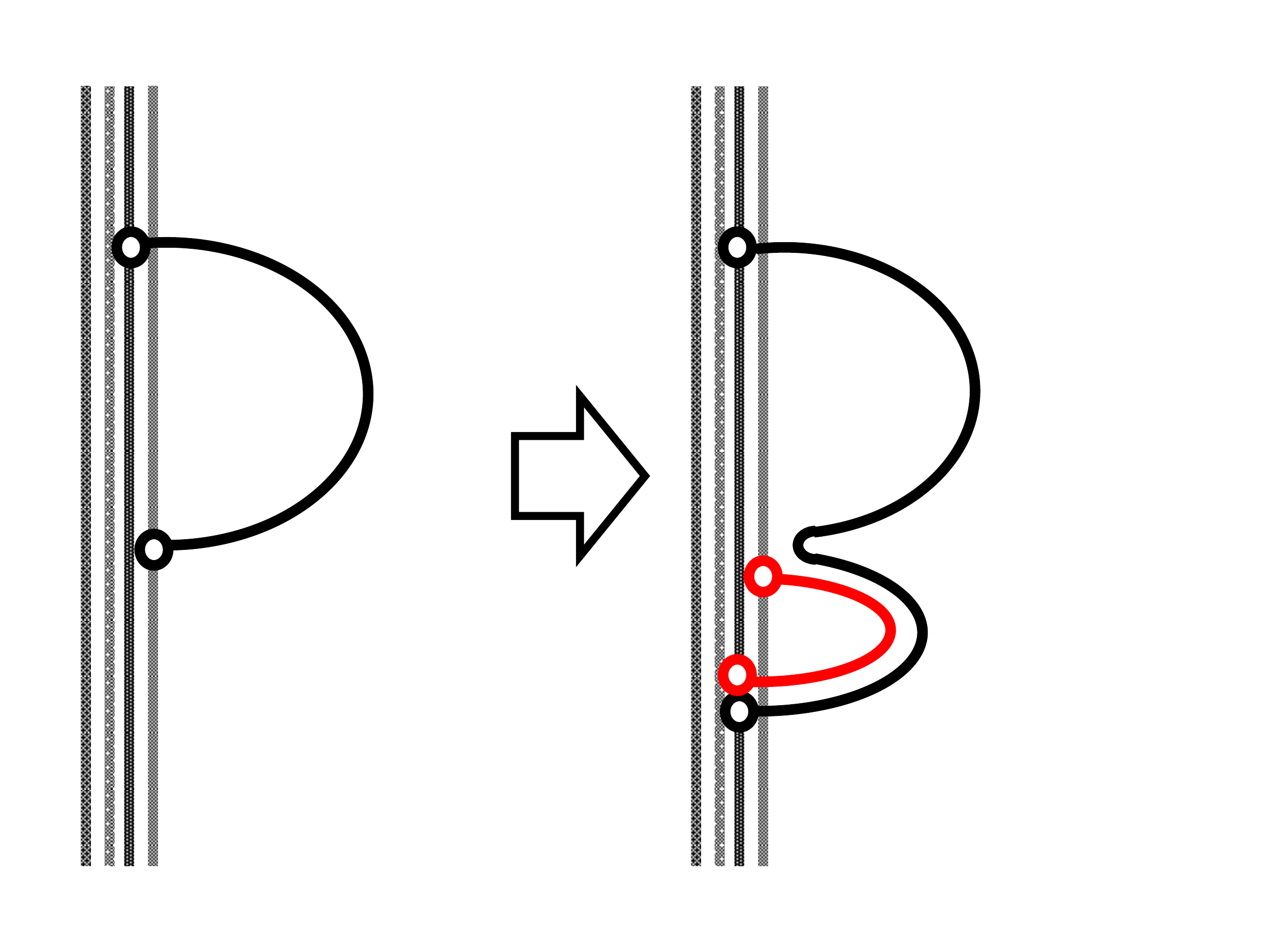}
   \\
(\ref{eq:FPH11})~~IK-type with\\
 open string $\bigcirc$
\end{center}
 \end{minipage}
\\
\vspace{5mm}
\\
$\bullet {\cal H}_3:$ Deformation on the line of open string
\\
 \begin{minipage}{0.24\hsize}
  \begin{center}
   \includegraphics[width=32mm]{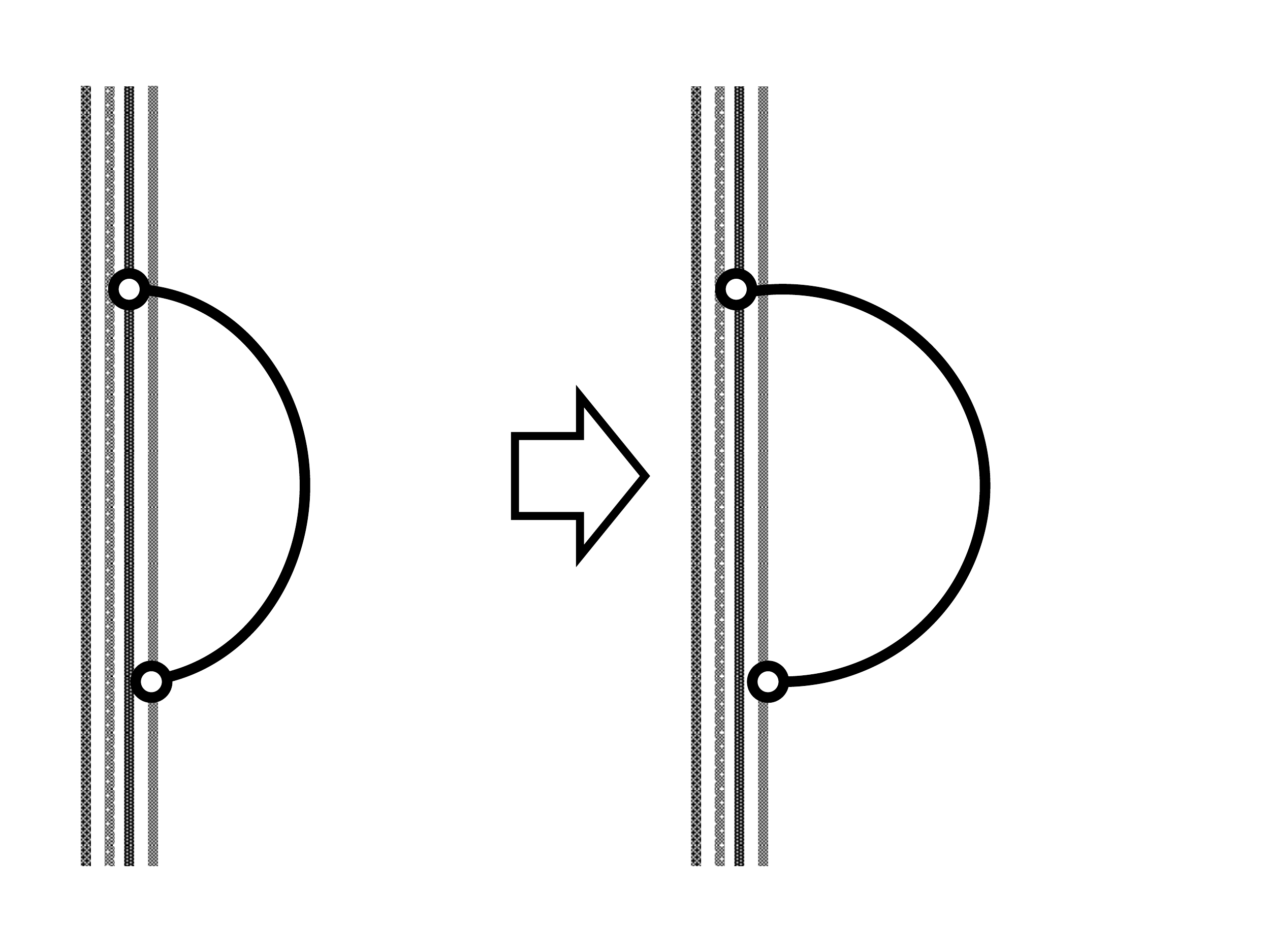}
   \\
(\ref{eq:FPH12})~~Propagation\\
 in a time $\times$
\end{center}
 \end{minipage}
 \begin{minipage}{0.24\hsize}
  \begin{center}
   \includegraphics[width=32mm]{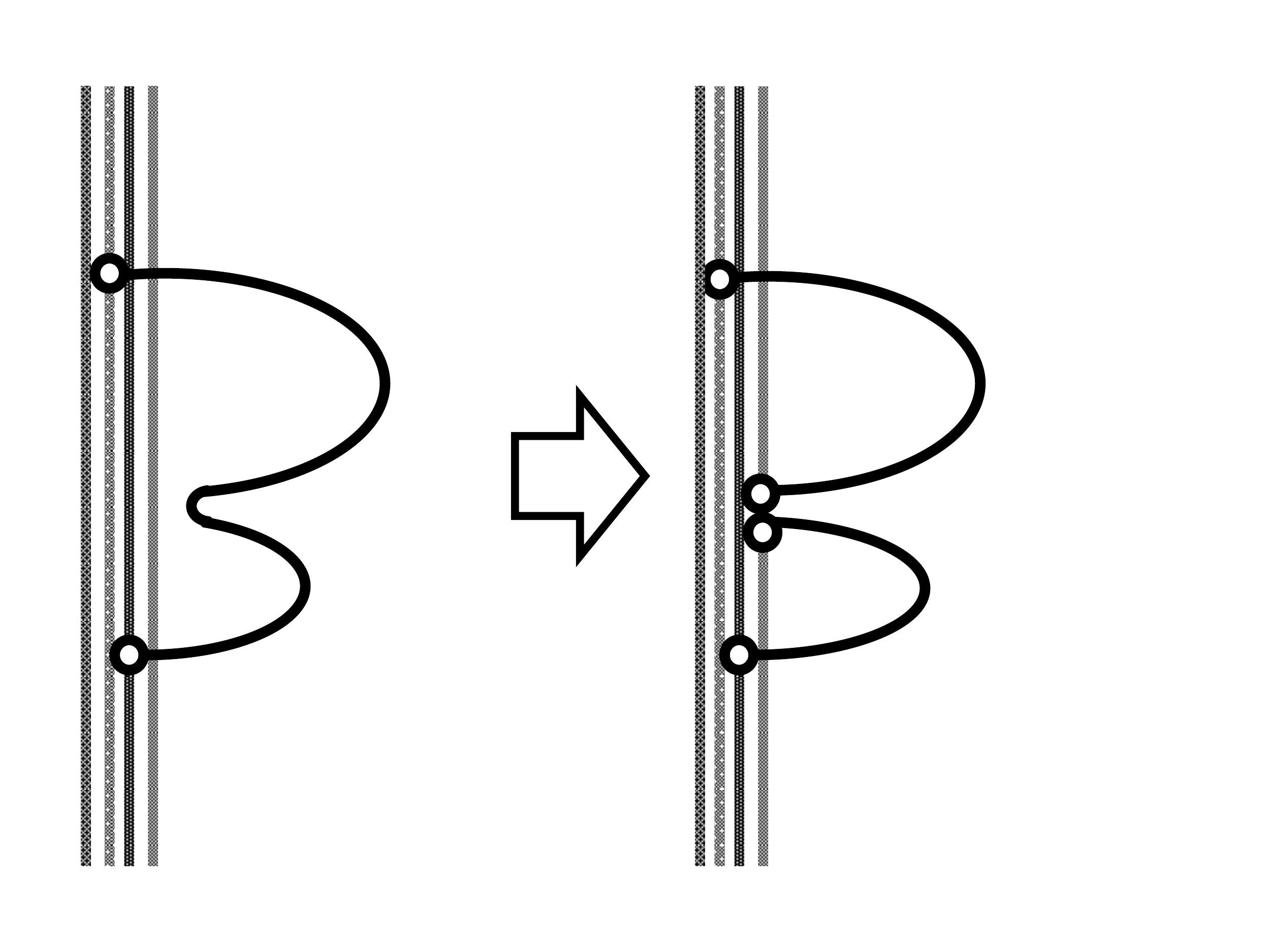}
   \\
(\ref{eq:FPH13})~~Splitting of\\
 open string $\times$
\end{center}
 \end{minipage}
 \begin{minipage}{0.24\hsize}
  \begin{center}
   \includegraphics[width=32mm]{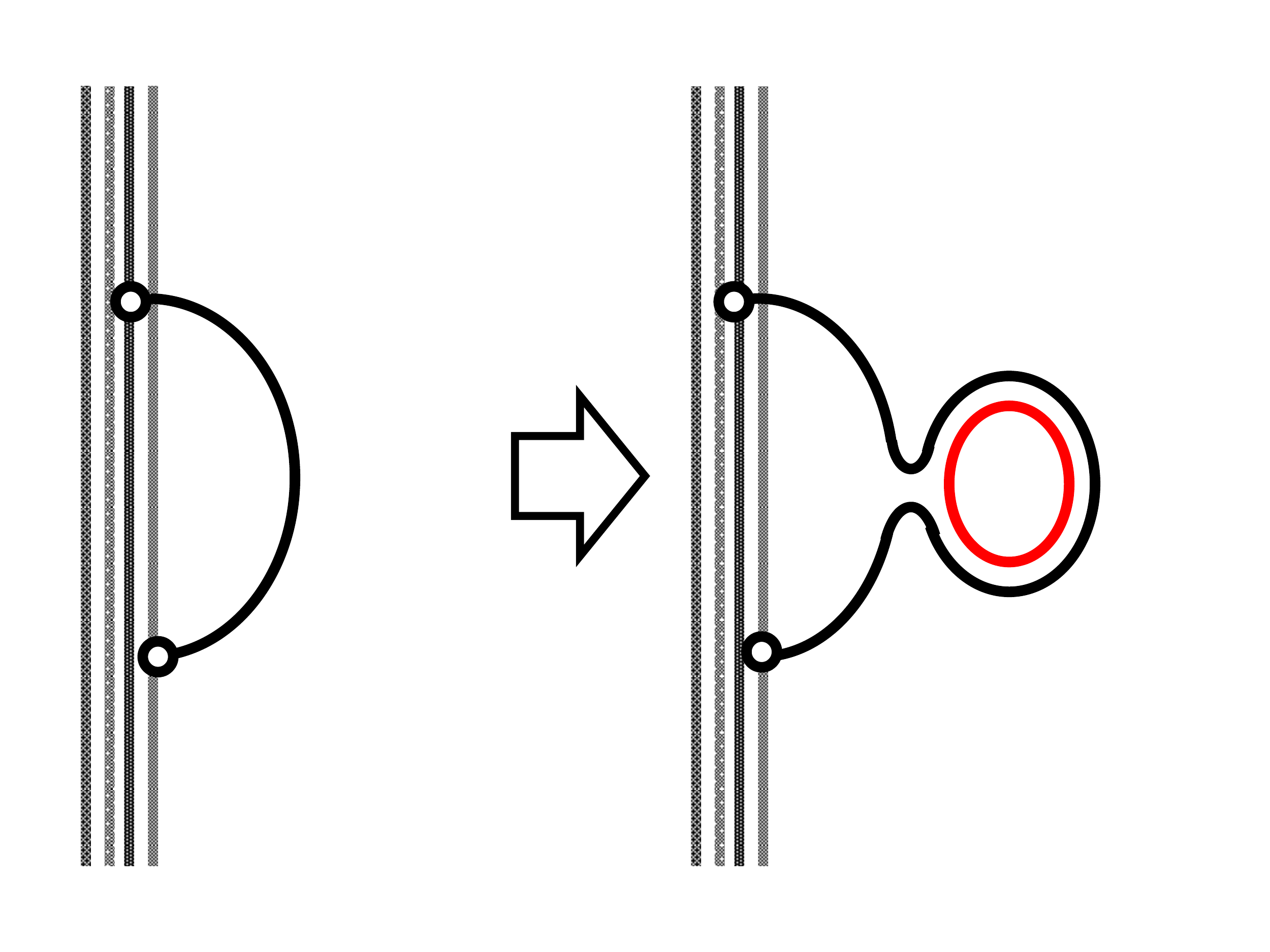}
   \\
(\ref{eq:FPH14})~~IK-type with\\
 closed string $\bigcirc$
\end{center}
 \end{minipage}
 \begin{minipage}{0.24\hsize}
  \begin{center}
   \includegraphics[width=32mm]{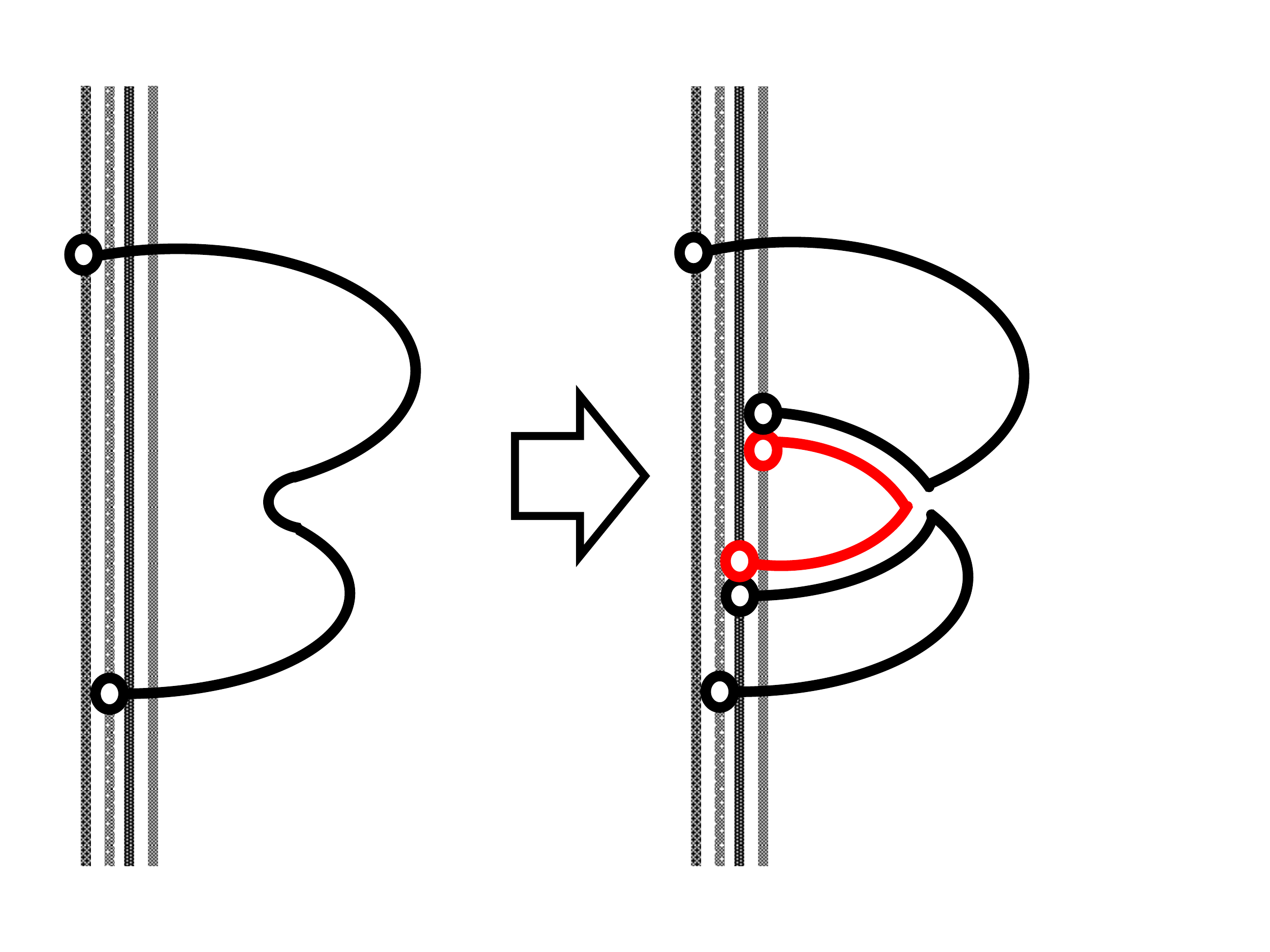}
   \\
(\ref{eq:FPH15})~~IK-type with\\
 open string $\bigcirc$
\end{center}
 \end{minipage}
\\
 \begin{minipage}{0.24\hsize}
  \begin{center}
   \includegraphics[width=32mm]{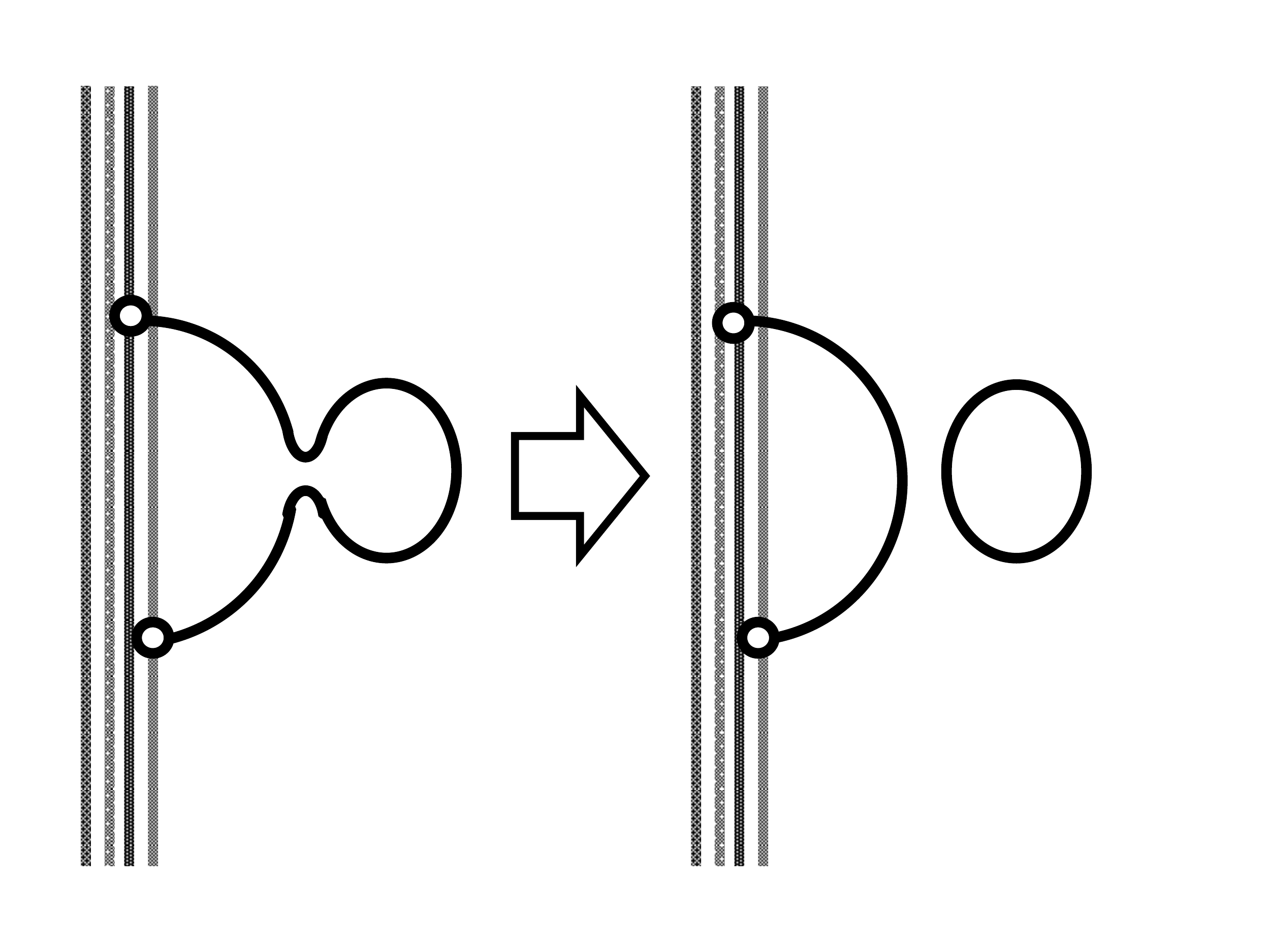}
   \\
(\ref{eq:FPH16})~~Splitting of\\
 closed string $\bigcirc$
\end{center}
 \end{minipage}
 \begin{minipage}{0.24\hsize}
  \begin{center}
   \includegraphics[width=32mm]{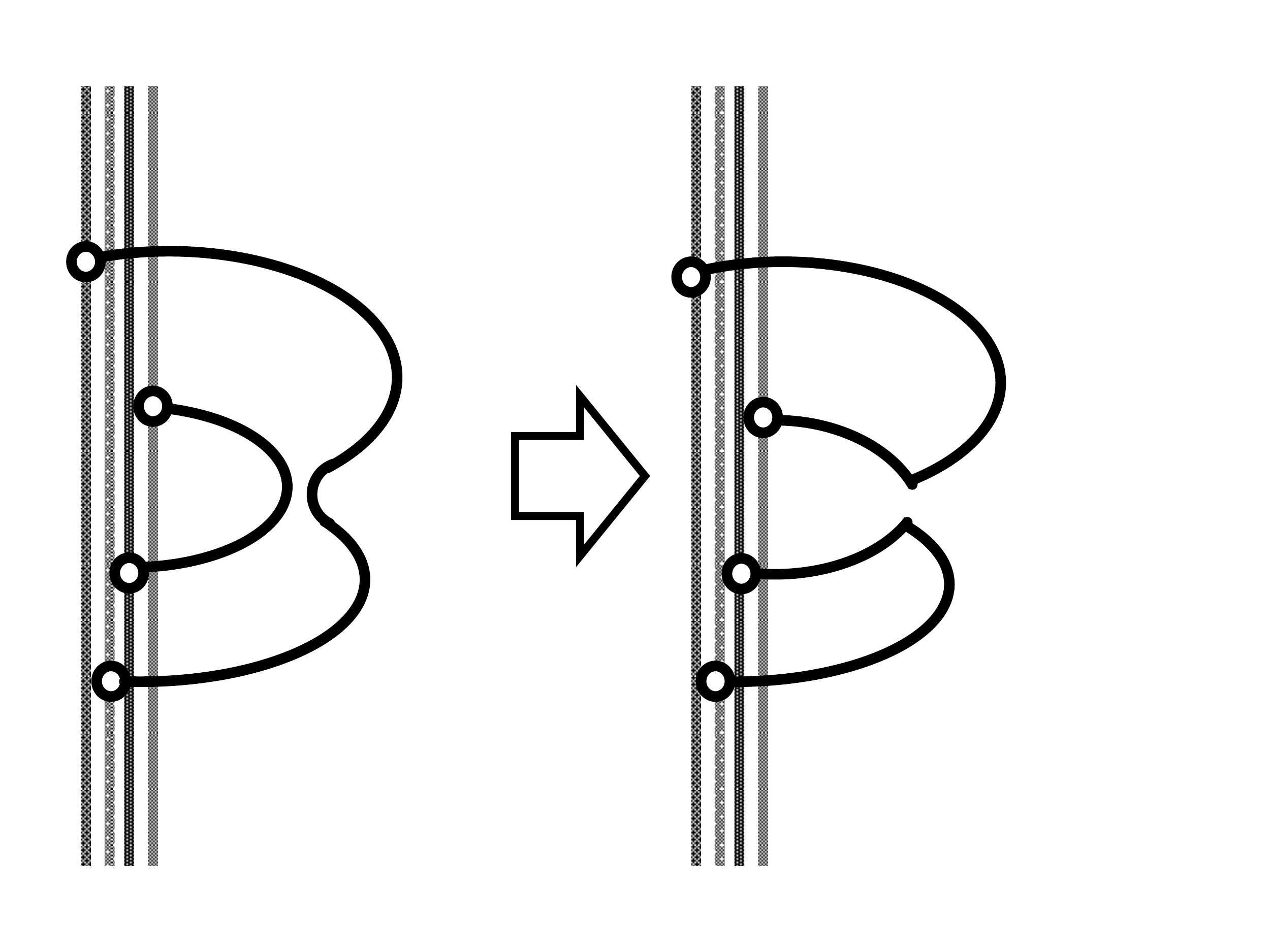}
   \\
(\ref{eq:FPH17})~~Cross-changing $\times$
\end{center}
 \end{minipage}
 \begin{minipage}{0.24\hsize}
  \begin{center}
   \includegraphics[width=32mm]{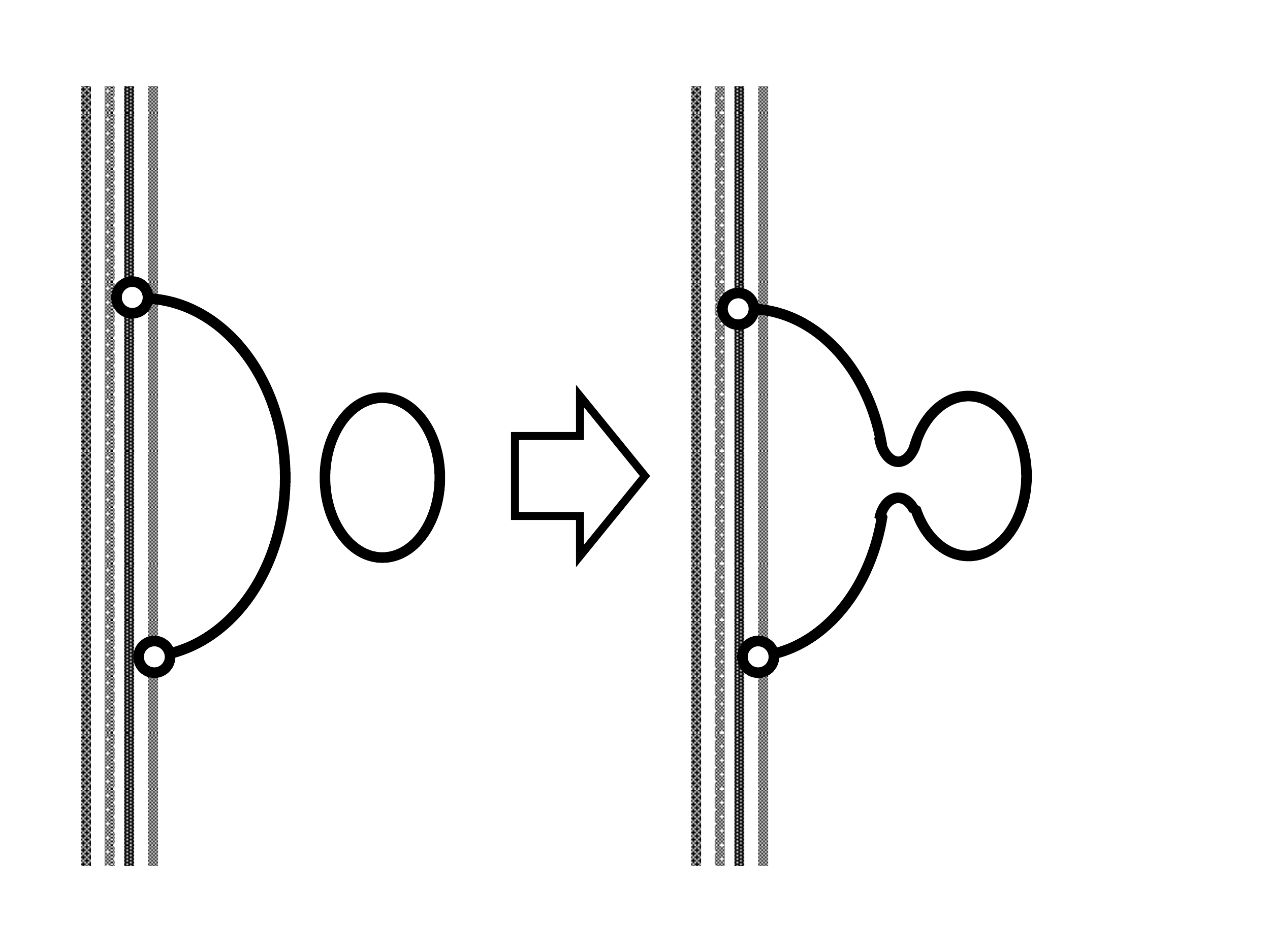}
   \\
(\ref{eq:FPH18})~~Merging with\\
 closed string $\times$
\end{center}
 \end{minipage}
\vspace{2mm}
\caption{
%18 processes of the stochastic time evolution of open-closed string:
The four straight lines express D-branes located at the same position.
The curved lines and the loops are open and closed strings, respectively.
The red lines (and loops) along the extended parts of the black lines (and loops) are strings created on the infinitesimal neighboring time by the IK-type interactions.
Interactions attached with ``$\bigcirc$" survive in ``the GCDT scaling'', while those with ``$\times$" scale out.
The interaction (\ref{eq:FPH7}) with ``$\triangle$'' is critical.
%Eqs.(\ref{eq:FPH6}) and (\ref{eq:FPH18}) are identical.
}
\end{figure}

Let us focus on ${\cal H}_1$, which is the processes for the closed string.
Four terms (\ref{eq:FPH1}), (\ref{eq:FPH3}),  (\ref{eq:FPH4}) and (\ref{eq:FPH5}) are exactly same ones with the GCDT model only of the closed string.
The scaling obtained in this previous model in ref.\cite{Kaw} was $4<D<6$ and $D_N < -D+1$, which we now call as ``the GCDT scaling''.
While the propagation in the equi-temporal slice (\ref{eq:FPH1}) and causality-violating merging interaction (\ref{eq:FPH4}) scale out, the splitting interaction (\ref{eq:FPH3}) and the IK-type interaction (\ref{eq:FPH5}) survive.
Three novel terms, (\ref{eq:FPH2}), (\ref{eq:FPH6}) and (\ref{eq:FPH7}), are interactions with the open string.
The merging interaction with an open string (\ref{eq:FPH6}), which explicitly breaks the causality, scales out in the GCDT scaling as we expect.
The term (\ref{eq:FPH2}) is interpreted as the merging interaction of the closed string with a D-brane, so it may break the causality.
Certainly it also scales out in this scaling.
The most interesting term is the IK-type interaction concerning the open strings (\ref{eq:FPH7}).
While for $D_N > -D$ this interaction becomes solely dominant, for $D_N < -D$ it scales out.
In the latter scaling, though the GCDT structure is kept, the closed string propagates and interacts just in the same way as the GCDT model only of closed string, or the closed string does not suffer any influence from the open string. 
Just when $D_N = -D$, we obtain more interesting model, in which the closed string propagator receives quantum correction by the interaction with D-branes.

The second part, ${\cal H}_2$, collects the processes on the edge ``$a$'' of the open string.
The term (\ref{eq:FPH8}) describes the open string propagation in the equi-temporal slice.
The scaling order becomes one order higher than that of the each original term because of the cancellation in the leading order.
This fact makes the open string propagation in the equi-temporal slice possible to scale out in the GCDT scaling, similarly to the term (\ref{eq:FPH1}) in ${\cal H}_1$.
In this scaling, the merging interaction (\ref{eq:FPH10}), that violates causality, becomes forbidden as it should.
We are left with (\ref{eq:FPH9}), connection of the edges of the open string to produce a closed string, and (\ref{eq:FPH11}), the IK-type interaction.

The third part, ${\cal H}_{2'}$ is same as ${\cal H}_2$, except that the deforming edge is ``$b$'' side.

The last part, ${\cal H}_3$, concerns the stochastic time evolution of an open string caused on some point except at the edges.
The term (\ref{eq:FPH12}) is the open string propagation in the equi-temporal slice, which becomes two orders higher than the original terms by the cancellation in the lowest two orders, so that it is managed to scale out just in the same way as the term of (\ref{eq:FPH1}).
The term (\ref{eq:FPH13}), the splitting of the open string into two open strings, scales out consistently, as it is the simmilar process to (\ref{eq:FPH2}).
Both of the terms (\ref{eq:FPH17}), the cross-changing of two open strings, and (\ref{eq:FPH18}), the merging interaction with a closed string, violate the causality and they scale out as we hope.
The remaining three interactions survive in this scaling as we expect from the analogy to the closed string model.
The IK-type interactions (\ref{eq:FPH14}) and (\ref{eq:FPH15}), concern a closed string creation and an open string creation, respectively, at the infinitesimal neighboring times. 
The term (\ref{eq:FPH16}) is the separation of a closed string from a open string with the total length conserved.
\begin{figure}[t]
 \begin{minipage}{0.47\hsize}
  \begin{center}
   \includegraphics[width=80mm]{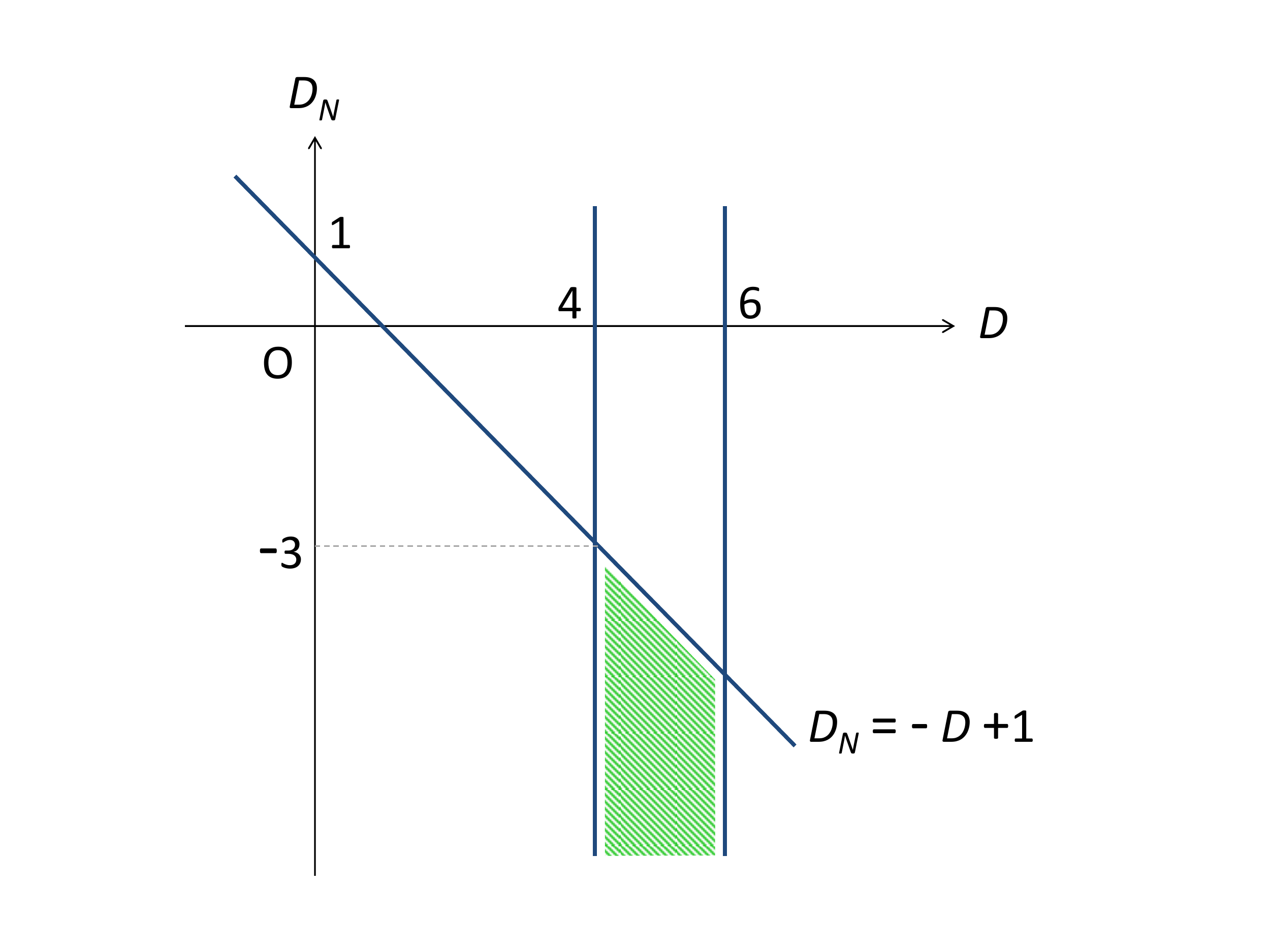}
\end{center}
\caption{``The GCDT scaling'' for closed string:
The closed string GCDT model is realized in the green area, which is classified into three phases in the open-closed string GCDT model.}
\label{fig:four}
 \end{minipage}
\hspace{5mm}
 \begin{minipage}{0.47\hsize}
  \begin{center}
   \includegraphics[width=80mm]{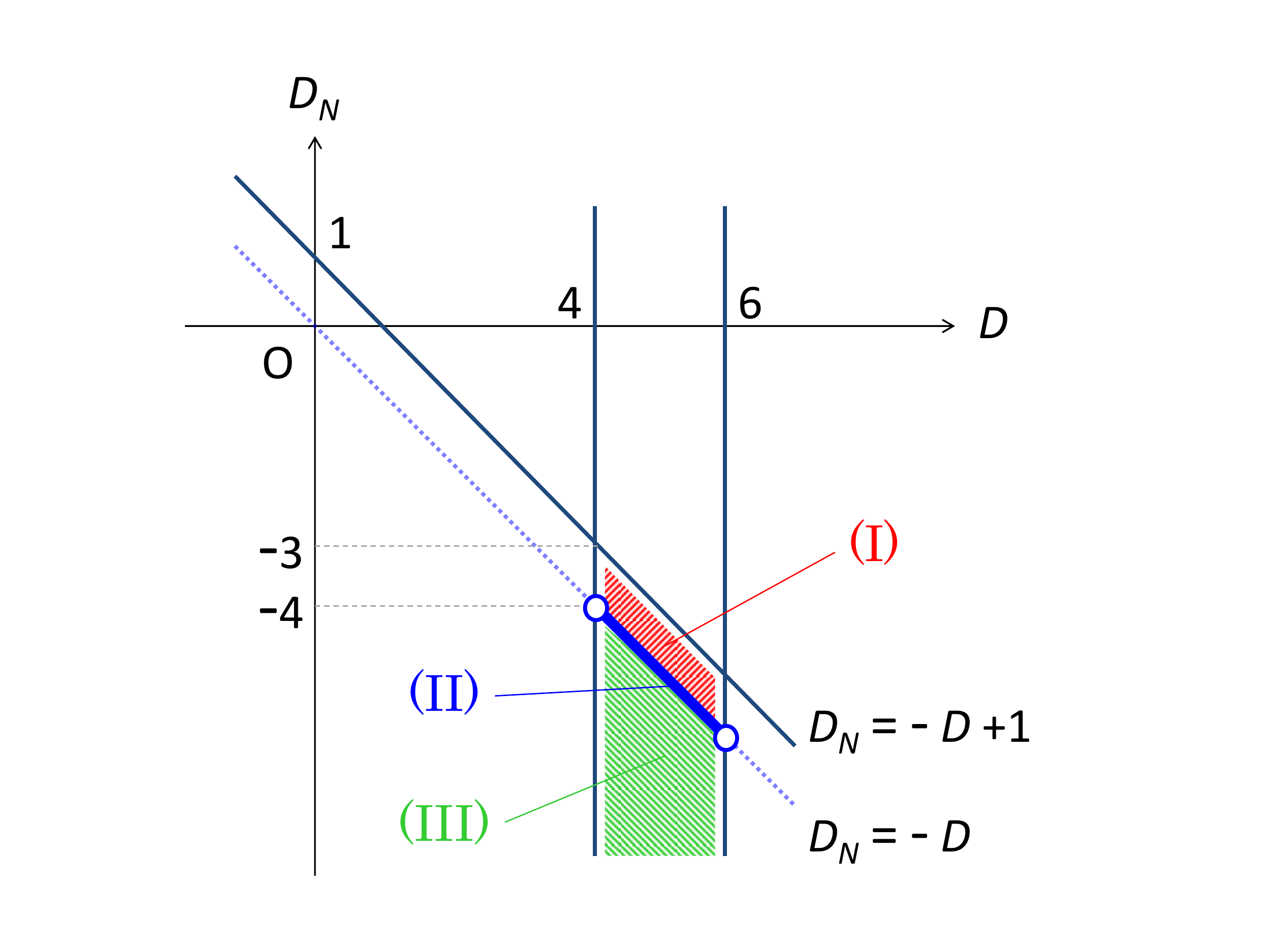}
\end{center}
\caption{Open-closed string GCDT:
In (I), the closed string is unstable.
In (III), the closed string suffers no influence by the open string.
On (II), the stable open-closed string model is obtained.
}
\label{fig:five}
 \end{minipage}

\end{figure}

\section{Conclusion}

We have constructed the matrix-vector model which realizes the CDT model of the open-closed string, as the extension of the CDT model of closed string.
Through the application of the stochastic quantization method, we obtain the GCDT model with the additional IK-type interactions, or the non-critical open-closed string field theory.
In this model, the stochastic time is not the geodesic distance any more but it is the step of the quantum correction.
The realization of the GCDT depends on two scaling dimensions, $D$ and $D_N$ (See Fig.\ref{fig:five}).
We obtain the restriction for $D$ as $4<D<6$, which is the same one with the closed string model.
Though in the closed string model the restriction for $D_N$ is only $D_N <-D +1$, in the open-closed string model we have three phases depending on the value of $D_N$.
(I) In the case $-D < D_N < -D+1$, the model is dominated only by the process of the string IK-type interaction (\ref{eq:FPH7}).
In this phase the closed string is unstable because any closed string tends to interact with D-branes so much that it becomes to open strings immediately.
(II) Just on $D_N =-D$, the open-closed string interacting model is realized, that is worth investing further.
(III) When $D_N <-D$, the processes of a closed string are independent of the existence of D-branes.
In this case, the processes directed from the open string to the closed string are irreversible.
In other words, the closed string model is inherited just as the subset of this open-closed string model. 
Therefore only in $D_N =-D$ we inspire D-branes with the physical substance.

\section*{Appendix}
In the appendix, we investigate the commutation relations of three generators, $L_t(n)$, $J_t^{ab}(n)$ and $K_t^{ab}(n)$, contained in the discrete F-P Hamiltonian, eq. (\ref{eq:DFPH}).
The expressions of the three generators are,
\footnote{
Eq. (\ref{eq:DFPH}) contains terms with $L_t(-1)$ and $K_t(-1)$.
We have to ignore the irrational splitting interaction terms in them, {\it i.e.} the second line in eq. (\ref{eq:generator1}) for $L_t(-1)$ and the last terms of the first and second lines in eq. (\ref{eq:generator3}) for $K_t^{ab}(-1)$.
}
\begin{eqnarray}
\label{eq:generator1}
L_t(n) & = & -N^2 \left[ g \phi _t(n+1) - \phi _t(n+2) +g \phi _t(n+3) + {1 \over N} \sum_c \psi _t^{cc}(n+1) \right. \nonumber \\
 & & + \sum_{k=0}^n \phi _t(k) \phi _t(n-k) \nonumber \\
 & & + \sum_{k=1}^{\infty} \phi _t(n+k) \left\{ {1 \over N^2} k \pi _t(k) + \hat{\phi}_t(k) \right\} \nonumber \\
 & & \left. + {1 \over N} \sum_{ab} \sum_{k=1}^{\infty} \psi _t^{ab}(n+k) k \left\{ {1 \over N} \pi _t^{ab}(k) + \hat{\psi}_t^{ba}(k) \right\} \right],  \\
\label{eq:generator2}
J_t^{ab}(n) & = & -N \left[ - \psi _t^{ab}(n) + g_B^a \psi _t^{ab}(n+1) + \delta ^{ab} \phi _t(n) \right. \nonumber \\ 
 & & \left. +\sum_c \sum_{k=0}^{\infty} \psi _t^{cb}(n+k) \left\{ {1 \over N}\pi _t^{ca}(k) + \hat{\psi}_t^{ac}(k) \right\} \right], \\
\label{eq:generator3}
K_t^{ab}(n) & = & - \left[ g \psi _t^{ab}(n+1) - \psi _t^{ab}(n+2) +g \psi _t^{ab}(n+3) +2 \sum_{k=0}^n \psi _t^{ab}(n-k) \phi _t(k) \right. \nonumber \\
 & & + \sum_c g_B^c \sum_{k=0}^{n+1} \psi _t^{ac}(n+1-k) \psi _t^{cb}(k) - \sum_c \sum_{k=0}^n \psi _t^{ac}(n-k) \psi _t^{cb}(k) \nonumber \\
 & & + {1 \over N} (n+1) \psi _t^{ab}(n) \nonumber \\
 & & + \sum_{k=1}^{\infty} \psi _t^{ab}(n+k) \left\{ {1 \over N^2} k \pi _t (k) + \hat{\phi}_t(k) \right\} \nonumber \\
 & & \left. + \sum_{cd} \sum_{k=0}^{\infty} \sum_{\ell =0}^{n+k} \psi _t^{ad}(\ell ) \psi _t (n+k-\ell ) \left\{ {1 \over N}\pi _t^{cd}(k) + \hat{\psi}_t^{dc}(k) \right\} \right].
\end{eqnarray}
They satisfy the following commutation relations:
\begin{eqnarray}
\left[ L_t(n), L_{t'}(m) \right] & = & (n-m) \delta _{tt'} L_t(n+m), \\
\left[ J_t^{ab}(n), J_{t'}^{cd}(m) \right] & = & \delta _{tt'} \delta ^{bc} J_t^{ad}(n+m) - \delta _{tt'} \delta ^{ad} J_t^{cb}(n+m), \\
\left[ L_t(n), J_{t'}^{ab}(m) \right] & = & -m \delta_{tt'}J^{ab}(n+m), \\
\left[ L_t(n), K_{t'}^{ab}(m) \right] & = & (n-m) \delta_{tt'}K_t^{ab}(n+m) \nonumber \\
 & & \! -{1 \over N} \delta _{tt'}  \! \sum_c \! \sum_{k=0}^{n-1} (n-k) \left\{ J_t^{cb}(n+m-k) \psi _t^{ac}(k) + J_t^{ac}(n+m-k) \psi _t^{cb}(k) \right\} \! , \nonumber \\
&& \\
\left[ J_t^{ab}(n), K_{t'}^{cd}(m) \right] & = & -\delta _{tt'} \delta ^{ad} K_t^{cb}(n+m) \nonumber \\
 & & +{1 \over N}\delta _{tt'} \delta^{ad} \sum_e \sum_{k=0}^{n-1} J_t^{ec*}(n+m-k) \psi _t^{eb}(k) \nonumber \\
 & & +{1 \over N}\delta _{tt'} \sum_{k=0}^{n-1} J_t^{ad}(n+m-k) \psi _t^{cb}(k), \\
\left[ K_t^{ab}(n), K_{t'}^{cd}(m) \right] & = & {1 \over N^2} \delta _{tt'} \sum_e \sum_{k=0}^{n-1} J_t^{ed}(n +m-k) \sum_{\ell =0}^k \psi _t^{ae}(\ell ) \psi _t^{cb}(k-\ell ) \nonumber \\
 & & + {1 \over N^2} \delta _{tt'} \sum_e \sum_{k=0}^{n-1} J_t^{ec*}(n +m-k) \sum_{\ell =0}^k \psi _t^{eb}(\ell ) \psi _t^{ad}(k-\ell ) \nonumber \\
 & & - {1 \over N^2} \delta _{tt'} \sum_e \sum_{k=0}^{m-1} J_t^{eb}(n +m-k) \sum_{\ell =0}^k \psi _t^{ce}(\ell ) \psi _t^{ad}(k-\ell ) \nonumber \\
 & & - {1 \over N^2} \delta _{tt'} \sum_e \sum_{k=0}^{m-1} J_t^{ea*}(n +m-k) \sum_{\ell =0}^k \psi _t^{ed}(\ell ) \psi _t^{cb}(k-\ell ).
\end{eqnarray}
The algebraic structure is the same type as that of the matrix-vector models for the non-critical string field theories\cite{AJ,EKN}.
Naively if we ignore the terms explicitly multiplied by $1/N$ and $1/N^2$, the commutators concerning $K_{t}^{ab}(n)$ look more familiar.
The first is the Virasoro algebra.
From the second relation, $J_t^{ab}(n) - J_t^{ba*}(n)$ is the generator of SU($R$) current algebra.

%%%%%%%%%%%%%%%%%%%%%%%%%%%%%%%%%%%%%%%%%%%%%%%%%%%%%%%%%%%%%%%%%%%%%%%%%%

%%%%%%%%%%%%%%%%%%%%%%%%%%%%%%%%%%%%%%%%%%%%%%%%%%%%%%%%%%%%%%%%%%%%%%%%%%%%%%%

\end{document}